\documentclass[a4paper,11pt]{article}
\pdfoutput=1
\RequirePackage{amsmath,amssymb}
\RequirePackage{graphicx}
\DeclareMathSymbol{\Ohm}{\mathalpha}{operators}{"0A}\providecommand*{\upOmega}{\Ohm}
\RequirePackage{siunitx}
\RequirePackage{booktabs}
\RequirePackage{url}
\usepackage{jinstpub}
\renewcommand{\eqref}[1]{eq.~(\ref{#1})}
\title{Technical design and commissioning of the KATRIN large-volume air coil system}
\author[a]{M.~Erhard,}
\author[a,b,1]{J.~Behrens,\note{Corresponding author.}}
\author[b]{S.~Bauer,}               
\author[c]{A.~Beglarian,}
\author[b]{R.~Berendes,}
\author[a,d]{G.~Drexlin,}
\author[d]{F.~Gl{\"u}ck,}
\author[d]{R.~Gumbsheimer,}
\author[d,x]{J.~Hergenhan,}         
\author[a]{B.~Leiber,}              
\author[d,e,f]{S.~Mertens,}
\author[g]{A.~Osipowicz,}
\author[d]{P.~Plischke,}
\author[d]{J.~Reich,}              
\author[d]{T.~Th{\"u}mmler,}
\author[d,y]{N.~Wandkowsky,}       
\author[b]{C.~Weinheimer}
\author[c]{and S.~W{\"u}stling}
\affiliation[a]{Experimental Particle Physics~(ETP), Karlsruhe Institute of Technology, Hermann-von-Helmholtz-Platz~1, D-76344 Eggenstein-Leopoldshafen, Germany}
\affiliation[b]{WWU M\"unster, Institut f\"ur Kernphysik, Wilhelm-Klemm-Str.~9, D-48149 M\"unster, Germany}
\affiliation[c]{Institute for Data Processing and Electronics~(IPE), Karlsruhe Institute of Technology, Hermann-von-Helmholtz-Platz~1, D-76344 Eggenstein-Leopoldshafen, Germany}
\affiliation[d]{Institute of Nuclear Physics~(IKP), Karlsruhe Institute of Technology, Hermann-von-Helmholtz-Platz~1, D-76344 Eggenstein-Leopoldshafen, Germany}
\affiliation[e]{Institute for Nuclear and Particle Astrophysics and Nuclear Science Division, Lawrence Berkeley National Laboratory, Berkeley, CA 94720, USA}
\affiliation[f]{Max-Planck-Institut f\"ur Physik, F\"ohringer Ring~6, 80805 M\"unchen, Germany}
\affiliation[g]{University of Applied Sciences~(HFD) Fulda, Leipziger Str.~123, 36037 Fulda, Germany}
\affiliation[x]{Now at: Institute for Anthropomatics and Robotics (IAR), Intelligent Process Automation and Robotics Lab (IPR), Karlsruhe Institute of Technology, Engler-Bunte-Ring~8, 76131 Karlsruhe, Germany}
\affiliation[y]{Now at: Wisconsin IceCube Particle Astrophysics Center~(WIPAC), 222 West Washington Ave., Suite 500 Madison, WI 53703, USA}
\emailAdd{jan.behrens@kit.edu}
\abstract{%
The KATRIN experiment is a next-generation direct neutrino mass experiment with a sensitivity of \SI{0.2}{eV} (90\%~C.L.) to the effective mass of the electron neutrino.
It measures the tritium $\beta$-decay spectrum close to its endpoint with a spectrometer based on the MAC-E filter technique.
The $\beta$-decay electrons are guided by a magnetic field that operates in the \si{mT} range in the central spectrometer volume; it is fine-tuned by a large-volume air coil system surrounding the spectrometer vessel.
The purpose of the system is to provide optimal transmission properties for signal electrons and to achieve efficient magnetic shielding against background.
In this paper we describe the technical design of the air coil system, including its mechanical and electrical properties.
We outline the importance of its versatile operation modes in background investigation and suppression techniques.
We compare magnetic field measurements in the inner spectrometer volume during system commissioning with corresponding simulations, which allows to verify the system's functionality in fine-tuning the magnetic field configuration.
This is of major importance for a successful neutrino mass measurement at KATRIN.

}
\begin{document}
\maketitle
\flushbottom
%
%
%
\section{Introduction}
\label{SecIntro}

The observation of neutrino-flavor oscillations has established solid evidence that neutrinos have finite mass~\cite{NuOsci1998}.
However, such measurements can only assess the squared mass difference of the three neutrino mass eigenstates, but not the absolute mass scale.
The neutrino mass is of major interest for models of mass generation mechanisms in the neutrino sector and has great cosmological relevance~\cite{Otten2008}.
In contrast to complementary studies of cosmological probes or neutrinoless double-$\beta$-decay, the kinematic measurement of the shape of the electron energy spectrum in $\beta$-decay is completely model-independent and relies only on energy-momentum conservation.

The \textbf{Ka}rlsruhe \textbf{Tri}tium \textbf{N}eutrino experiment (KATRIN)~\cite{KATRIN2005} is designed to measure the `effective mass of the electron neutrino', given by an incoherent sum over the mass eigenstates~\cite{Drexlin2013}, with a sensitivity down to \SI{0.2}{eV} (90\%~C.L.).
Since only a small fraction of the emitted electrons near the spectrum endpoint carries information about the neutrino mass, a high signal rate and low background in the endpoint region is essential.
KATRIN uses tritium with a low endpoint $E_0(\text{T}_2) = \SI{18571.8(12)}{eV}$~\cite{Otten2008,Myers2015} and a short half-life $t_{1/2} = \SI{12.32(2)}{a}$~\cite{TritiumHalflife2000} as a nearly ideal $\beta$-emitter for this type of experiment~\cite{Drexlin2013}.

\begin{figure}[tb]
    \includegraphics[width=\columnwidth]{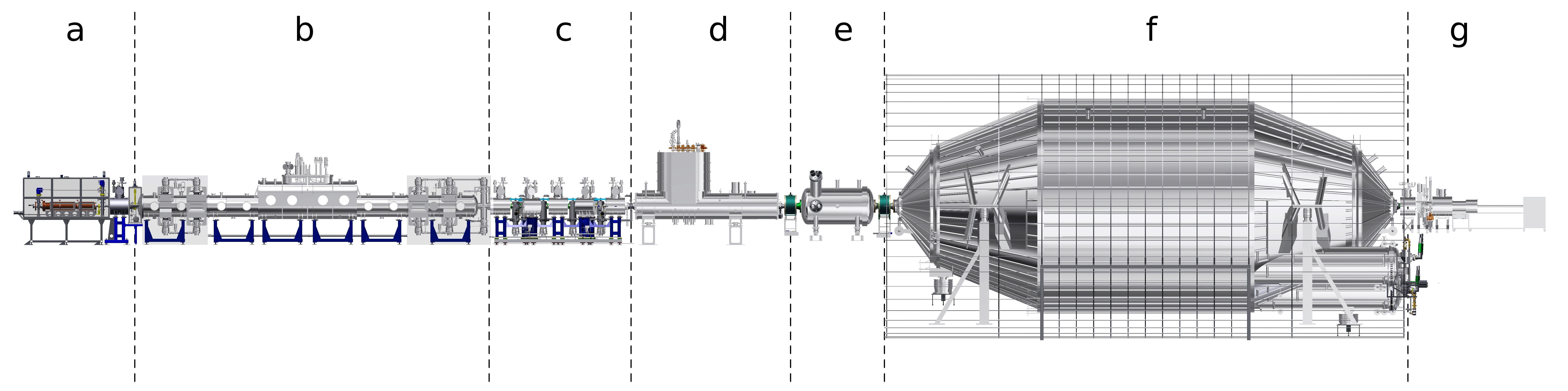}
    \caption{Beamline of the KATRIN experiment.
        Electrons are emitted from $\beta$-decay in (b) the windowless gaseous tritium source (WGTS); the rear section (a) contains calibration devices and closes the source end of the setup.
        Following the source is the transport section, consisting of (c) the differential pumping section (DPS) and (d) the cryogenic pumping section (CPS).
        The spectrometer and detector section consists of (e) the pre-spectrometer, (f) the main spectrometer with the large-volume air coil system and (g) the focal-plane detector (FPD).
        }
    \label{FigKATRIN}
\end{figure}

A schematic overview of the \SI{70}{m} long KATRIN setup is shown in figure~\ref{FigKATRIN}.
Molecular gaseous tritium of high purity is injected into the windowless source (WGTS), where a small fraction of the tritium molecules decays inside the \SI{10}{m} long beam tube that is operated at about \SI{30}{K}~\cite{Priester2015}.
Superconducting magnets at the source and the transport section produce magnetic fields up to \SI{5.6}{T}, which adiabatically guide all forward-emitted $\beta$-decay electrons towards the spectrometer and detector section.
In the transport section, tritium molecules are removed from the beam line by a sequence of turbo-molecular pumps (TMPs) at the differential pumping system (DPS)~\cite{Lukic2012} and by cryosorption on a \SI{3}{K} surface covered in argon frost at the cryogenic pumping section (CPS)~\cite{Gil2010}.
The magnets and beam tube in the pumping sections form a chicane to block the direct line of sight from source to spectrometers, in order to prohibit a molecular beaming effect. This retention system achieves a tritium suppression by 14~orders of magnitude~\cite{Luo2008}.
Charged particles are guided around the chicane by magnetic field lines; positive ions from $\beta$-decays in the WGTS are therefore electrostatically blocked in the transport section, so that only electrons from tritium $\beta$-decays enter the spectrometer section.

The kinetic energy of the signal electrons is analyzed by a spectrometer that is based on the principle of magnetic adiabatic collimation with electrostatic filtering (MAC-E filter)~\cite{Lobashev1985,Picard1992}, where a retarding potential is applied to reflect electrons below a given energy threshold.
This key technique is explained in more detail in section~\ref{sec:MACE}.
Electrons with energies far below the endpoint carry no information about the neutrino mass.
A smaller MAC-E filter (the pre-spectrometer) rejects these electrons, thereby reducing the number of electrons entering the main spectrometer by 6~orders of magnitude~\cite{KATRIN2005}.
In the main spectrometer, all electrons with sufficient kinetic energy to pass the MAC-E filter are guided to the focal-plane detector (FPD) system~\cite{Amsbaugh2015} where they are counted.
The $\beta$-spectrum close to $E_0$ is measured in integral mode by variation of the main spectrometer retarding potential.
The main spectrometer is operated at ultra-high vacuum $p \le \SI{e-10}{mbar}$ in order to reduce scattering processes of electrons with residual gas~\cite{Arenz2016}.

This paper focuses on the technical realization of the air coil systems that fine-tune the shape of the magnetic field in the KATRIN main spectrometer~\cite{Glueck2013}.
The outline of this paper is as follows:
In section~\ref{SecEMD} we briefly list the general design criteria of the air coil systems.
In section~\ref{SecTD} we discuss the electrical design of the air coil systems and the mechanical design of its support structure.
Section~\ref{SecPerformance} gives an overview of the measurements that have been performed to confirm the functionality of the air coil systems, to characterize the magnetic field in the main spectrometer and to verify corresponding magnetic field simulations.
In particular we include a direct comparison of measured and simulated fields, as well as investigations on the linearity of the air coil currents w.r.t. to the achieved magnetic field in the spectrometer.

\section{Electromagnetic Design}
\label{SecEMD}

The main spectrometer is the central KATRIN component to analyze the kinetic energy of electrons from tritium $\beta$-decay.
The energy analysis is performed by applying a precision electrostatic retarding potential, which reflects all electrons with kinetic energies below the so-called transmission threshold.
The spectrometer is designed as a MAC-E filter and thus acts as an integral high-pass energy filter with large angular acceptance.

\subsection{The MAC-E filter technique}
\label{sec:MACE}

The principle of the MAC-E filter achieves a collimation of isotropically emitted electrons by the inverse magnetic mirror effect~\cite{Beamson1980,Lobashev1985,Picard1992}.
The technique provides superior energy resolution in the \si{eV} range at electron energies of several \si{keV} and a high luminosity for signal electrons.
It is thus an excellent choice for the KATRIN experiment~\cite{Drexlin2013}.

In case of adiabatic electron propagation, the orbital magnetic moment $\mu$ of the electron's cyclotron motion around a magnetic field line is conserved,
\begin{equation}
    \label{eq:magneticmoment}
    \mu = \frac{E_\perp}{B} = \text{const.}
    \,,
\end{equation}
in a non-relativistic approximation. Here $E_\perp$ and $B$ denote the transversal electron energy and the magnetic field, respectively.
The transversal kinetic energy can be written in terms of the kinetic energy $E$ and the pitch angle $\theta = \angle(\vec{p}, \vec{B})$ between electron momentum $\vec{p}$ and magnetic field $\vec{B}$, so that $E_\perp = E \cdot \sin^2\theta$.

Signal electrons enter the spectrometer at a magnetic field $B_\mathrm{max}$ that is produced by a superconducting solenoid at the spectrometer entrance; in a symmetric setup another solenoid at the spectrometer exit operates at the same magnetic field\footnote{
    KATRIN typically uses an asymmetric setup with a higher magnetic field at the exit}.
A magnetic field gradient towards the spectrometer center leads to a decrease of the transversal kinetic energy $E_\perp$ due to the condition \eqref{eq:magneticmoment}.
The transversal kinetic energy reaches its minimum in the so-called analyzing plane, where the magnetic field reaches a small value $B_\mathrm{a} \ll B_\mathrm{max}$ in the \SI{0.5}{mT} range.
Concurrently, the transversal energy is transformed into longitudinal energy due to energy conservation, which is then analyzed by the maximum retarding potential $|U_0|$ in the analyzing plane.

Electrons are transmitted if their longitudinal kinetic energy is larger than the filter energy, $E_\parallel = E \cdot \cos^2\theta > q U_0$, where $q$ is the electron charge.
The energy resolution achieved by an MAC-E filter is given by the ratio~\cite{Otten2008}
\begin{equation}
    \label{eq:energyresolution}
    \Delta E = E \cdot \frac{B_\mathrm{a}}{B_\mathrm{max}}
    \,.
\end{equation}
The KATRIN setup achieves an unprecedented energy resolution of $\Delta E = E \cdot \SI{0.3}{mT} / \SI{6}{T} \approx \SI{1}{eV}$ at $E = \SI{18.6}{keV} \approx E_0$.

\subsection{Electromagnetic design requirements}
\label{sec:EMD}

A fundamental requirement for the MAC-E filter technique is to guarantee a strictly adiabatic propagation of signal electrons.
This is ensured by maintaining sufficiently small magnetic field gradients and by initiating electrostatic retardation already in the high-field region near the spectrometer entrance and exit.
In case of small field gradients, the momentum transformation \eqref{eq:magneticmoment} occurs over a distance of several meters, resulting in an overall length of \SI{23}{m} of the KATRIN main spectrometer.

At KATRIN, a magnetic flux tube tube of $\Phi = \SI{191}{T.cm^2}$ transports signal electrons through the beam line and the spectrometers to the detector without any interference by mechanical structures through the spectrometers.
The outer radius $r_\mathrm{max}$ of the flux tube is defined by the conservation of magnetic flux~\cite{KATRIN2005},
\begin{equation}
    \label{eq:fluxtube}
    \Phi = \int \vec{B} \,\mathrm{d}\vec{A} \approx B \cdot \pi r_\mathrm{max}^2 = \text{const.}
    \,,
\end{equation}
where $B$ is the magnetic field at a given position along the beamline and $\vec{A}$ is the flux tube cross-section.

It is not possible to confine the entire flux tube inside the \SI{10}{m} diameter main spectrometer vessel solely by the stray field of the superconducting solenoids (see fig.~3 in \cite{Glueck2013}).
Furthermore, the magnetic field inside the main spectrometer is distorted due to the influence of the earth magnetic field and other contributions (sec.~\ref{SecPerformance}).
If the magnetic flux tube penetrates mechanical structures, a loss of signal electrons and an increase in background occurs since electrons from secondary emission at the metal surfaces of the spectrometer are transported directly to the detector.
Both issues can be resolved by implementing a large-volume air coil system around the main spectrometer that fine-tunes the magnetic field and ensures that the flux tube fits inside the spectrometer vessel.

The requirements on the magnetic field for a MAC-E filter and the electromagnetic design of the KATRIN air coil system are given in~\cite{Glueck2013} and a detailed discussion is available in~\cite{PhDWandkowsky2013}.
Here we briefly summarize the design criteria:
\begin{itemize}
    \item Transmission properties:
        The electromagnetic configuration must ensure that the transformation of transversal into longitudinal energy of an electron is completed before it reaches the maximum electrostatic retarding potential.
        The magnetic field generated by the air coil systems can be fine-tuned to fulfill this requirement.

    \item Energy resolution:
        The magnetic field $B_\mathrm{a}$ in the analyzing plane defines the energy resolution of the spectrometer for a given $B_\mathrm{max}$ and is a key parameter for neutrino mass analysis.
        Due to the limited spatial resolution of the pixelated detector wafer~\cite{Amsbaugh2015}, a nearly homogeneous magnetic field is required to minimize systematic uncertainties.

    \item Background reduction:
        Background is induced by secondary electrons created from electron emission at the inner spectrometer surfaces.
        The dominant mechanism to reduce this background is magnetic shielding by the Lorentz force~\cite{PhDWandkowsky2013}.
        This magnetic shielding, however, requires an excellent axial symmetry of the electromagnetic fields~\cite{Glueck2005}; otherwise surface-generated electrons would quickly drift into the inner flux tube volume and cause background.
        The dominant non-axial magnetic field component is the earth magnetic field; other sources include stray fields from magnetized steel bars and rods in the building materials of the spectrometer hall.
\end{itemize}

In order to fulfill the these electromagnetic design requirements, two independent (but complementary) large-volume air coil systems surround the KATRIN main spectrometer: the earth magnetic-field compensation system (EMCS) and the axial low-field correction system (LFCS); the LFCS allows to fine-tune the shape and strength of the magnetic field in the main spectrometer.
The technical realization of these systems (sec.~\ref{SecTD}) and corresponding performance tests during system commissioning (sec.~\ref{SecPerformance}) are detailed below.

\subsection{Simulation of the electromagnetic design}
\label{sec:KASSIOPEIA}

During any measurement phase, the magnetic field values in the flux tube can only be assessed and monitored by precision measurements outside the spectrometer and by precise modeling of all external magnetic field components.
The used magnetic field model must be validated by extensive test campaigns in order to achieve the desired accuracy.

For the initial design of the axially-symmetric LFCS, the \textsc{PartOpt} code\footnote{\url{http://www.partopt.net}} was employed.
It uses elliptic integrals for magnetic field calculations of axially symmetric coils.
In contrast to the elliptic integral method, the zonal harmonic expansion method can be \numrange{100}{1000} times faster and also provides a more general solution than the similar radial series expansion~\cite{Glueck2011b}.
Its major benefits are fast computation and high accuracy, which makes the method well-suited especially for trajectory calculations of charged particles in electromagnetic fields as in the case of KATRIN.
The original field calculation codes have been rewritten into a C++ framework~\cite{PhDCorona2014,PhDFurse2015} and were included into the \textsc{Kassiopeia}~\cite{Furse2017} software that is now the standard simulation package\footnote{\url{https://github.com/KATRIN-Experiment/Kassiopeia} (public release)} for the KATRIN experiment.

The magnetic field of the linear current sections of the EMCS is computed by integrating the Biot-Savart formula~\cite{PhDErhard2016}.
The curved arcs of the wire loops are approximated by short linear current segments.
Magnetic materials in the main spectrometer building are modeled by an ensemble of dipole bars~\cite{PhDReich2013}.
The contribution of structural materials to the magnetic field can be calculated if their magnetization is known; this is further discussed in sec.~\ref{SecPerformance}.

\section{Technical Design of the Air Coil Systems}
\label{SecTD}

\begin{figure}[tb]
    \centering
    \includegraphics[width=0.90\textwidth]{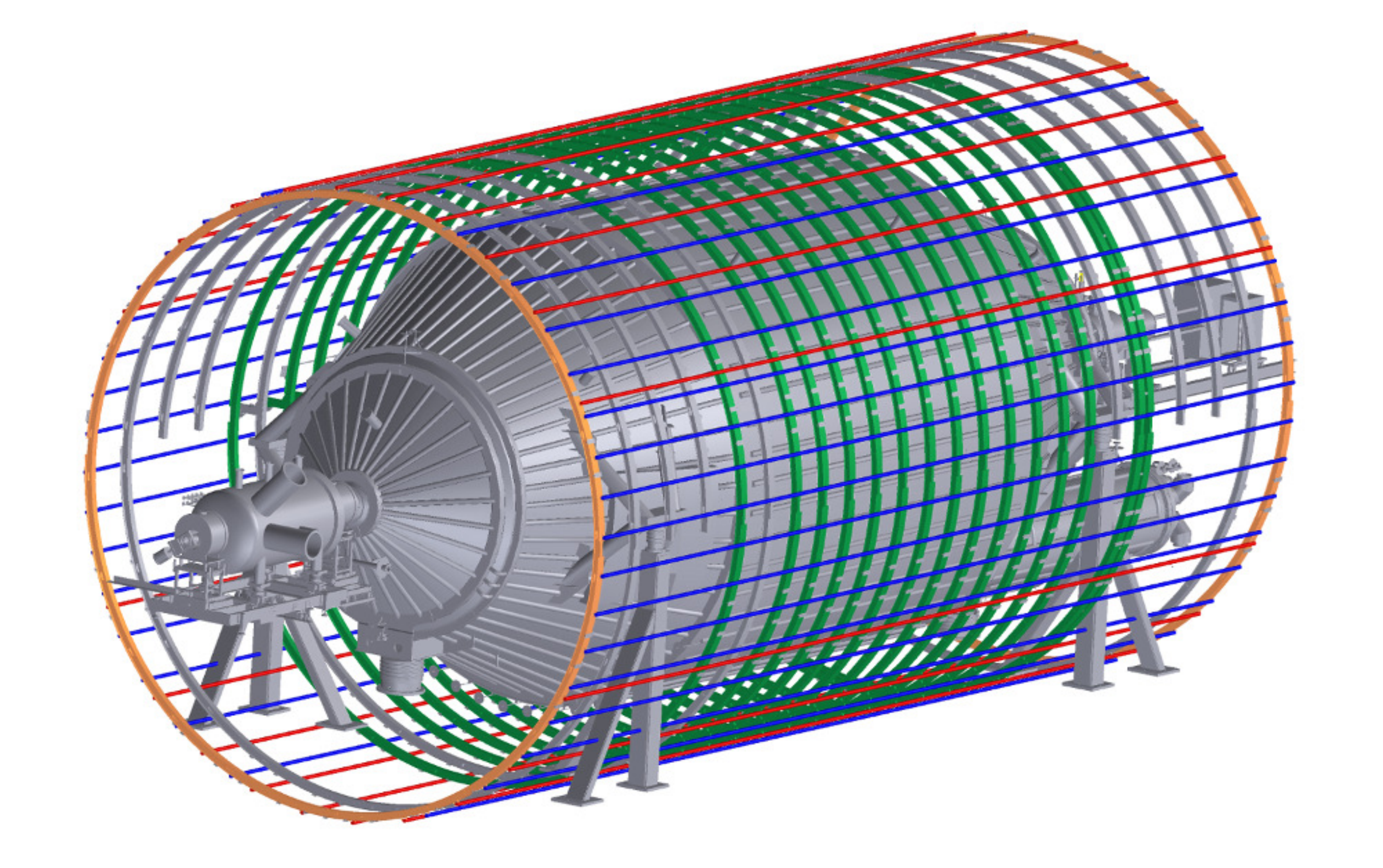}
    \caption{The KATRIN main spectrometer is surrounded by two large-volume air coil systems: the low field correction system (LFCS) and the earth magnetic field compensation system (EMCS) are both mounted on a cylindrical holding structure.
        The green circles represent the LFCS coils; the blue and red lines correspond to the current loops of the vertical and the horizontal EMCS components, respectively.
        The orange circles at the ends contain the arc segments that connect the linear current segments of the EMCS.
    }
    \label{FigMainspec}
\end{figure}

An overview of the large volume air coil systems as obtained from CAD drawings is shown in fig.~\ref{FigMainspec}.
This section presents the technical design of these systems, starting with an overview of the system layout (sec.~\ref{sec:layout_overview}) and continuing with details regarding the mechanical (sec.~\ref{sec:mechanical}) and electrical (sec.~\ref{sec:electrical}) design.
Finally, the influence of deformations of the air coil system due to its significant mass and the effect on the magnetic field is discussed in sec.~\ref{sec:gravity}.

\subsection{LFCS System Layout}
\label{sec:layout_overview}

\begin{figure}[tb]
    \centering
    \includegraphics[width=1.00\textwidth]{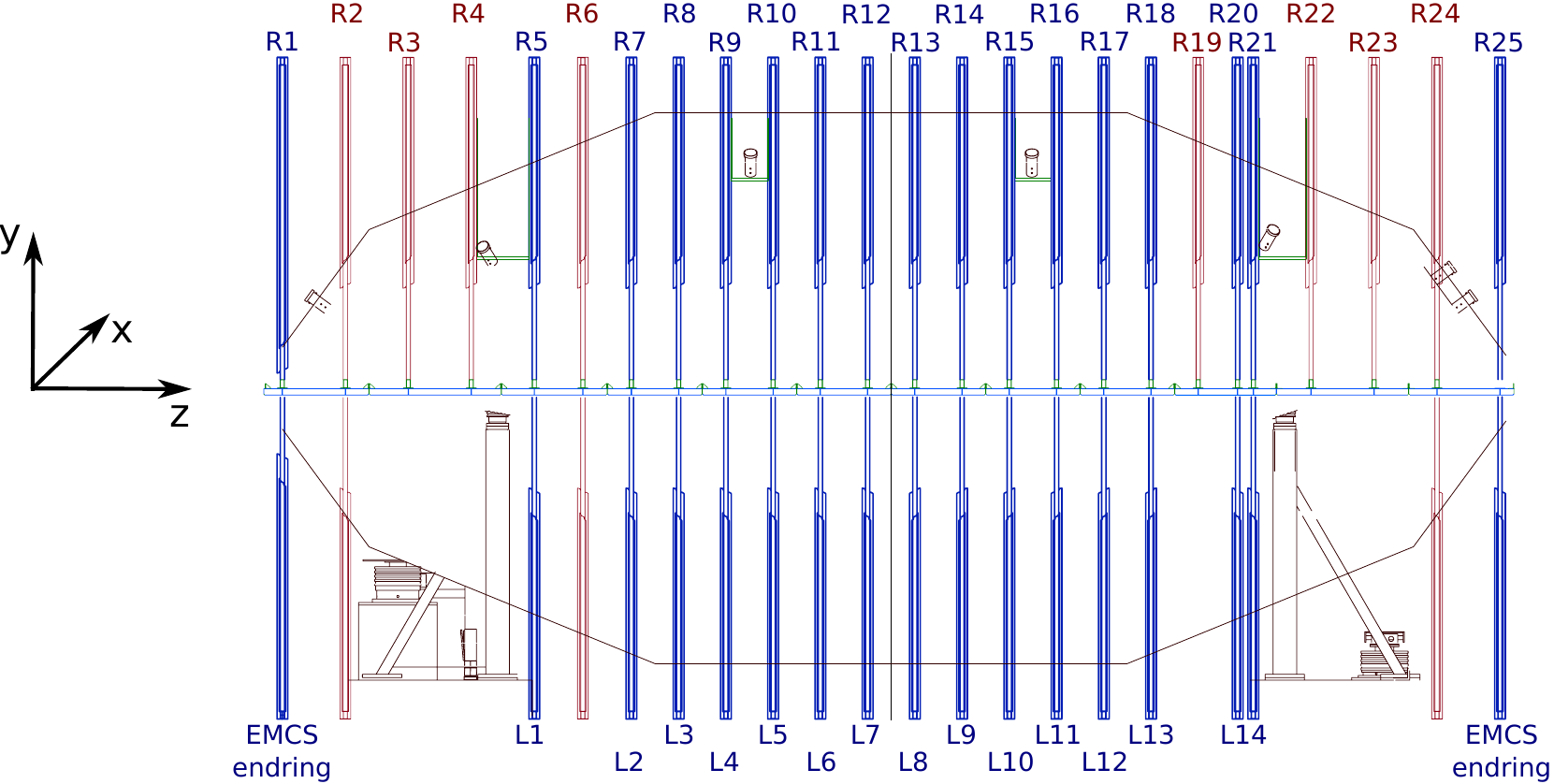}
    \caption{The large-volume air coil system consists of 25 axially aligned rings with a diameter of \SI{12.6}{m} that are labeled $\text{R1--R25}$.
        The central coils $\text{R5}, \text{R7--R18}, \text{R20+R21}$ constitute the LFCS and are indicated by coil numbers $\text{L1--L14}$ while the end rings $\text{R1}, \text{R25}$ are used for the EMCS; both groups are highlighted in blue.
        The remaining rings highlighted in red are required for stability of the entire system; they are also used to guide the EMCS wire loops (not shown here, see fig.~\ref{FigMainspec}).
    }
    \label{fig:LFCS}
\end{figure}

The LFCS, which is detailed in fig.~\ref{fig:LFCS}, consists of 14 air coils that are coaxially aligned with the spectrometer axis (beam axis) and the two superconducting solenoids located at the entrance (the PS2 magnet at \SI{4.5}{T}) and exit (the PCH magnet at \SI{6}{T}) of the main spectrometer.
Each air coil has a diameter of \SI{12.6}{m} and is powered by a separate power supply (details in sec.~\ref{sec:electrical}).
With this coil arrangement it is possible to precisely tune the magnetic field to the desired shape and strength, which is necessary to fulfill the magnetic field requirements discussed above.
The air coil system is designed to generate magnetic fields of up to \SI{1.8}{mT} in the center of the spectrometer.
Combined with the solenoid stray field contribution of \SI{0.2}{mT}, a total field strength of \SI{2.0}{mT} in the analyzing plane can be maintained.
For neutrino mass measurements, the magnetic field needs to be larger than \SI{0.33}{mT} in order to fit the entire flux tube of \SI{191}{T.cm^2} into the spectrometer vessel according to \eqref{eq:fluxtube}.

\subsection{EMCS system layout}

\begin{figure}[tb]
    \centering
    \includegraphics[height=5.5cm]{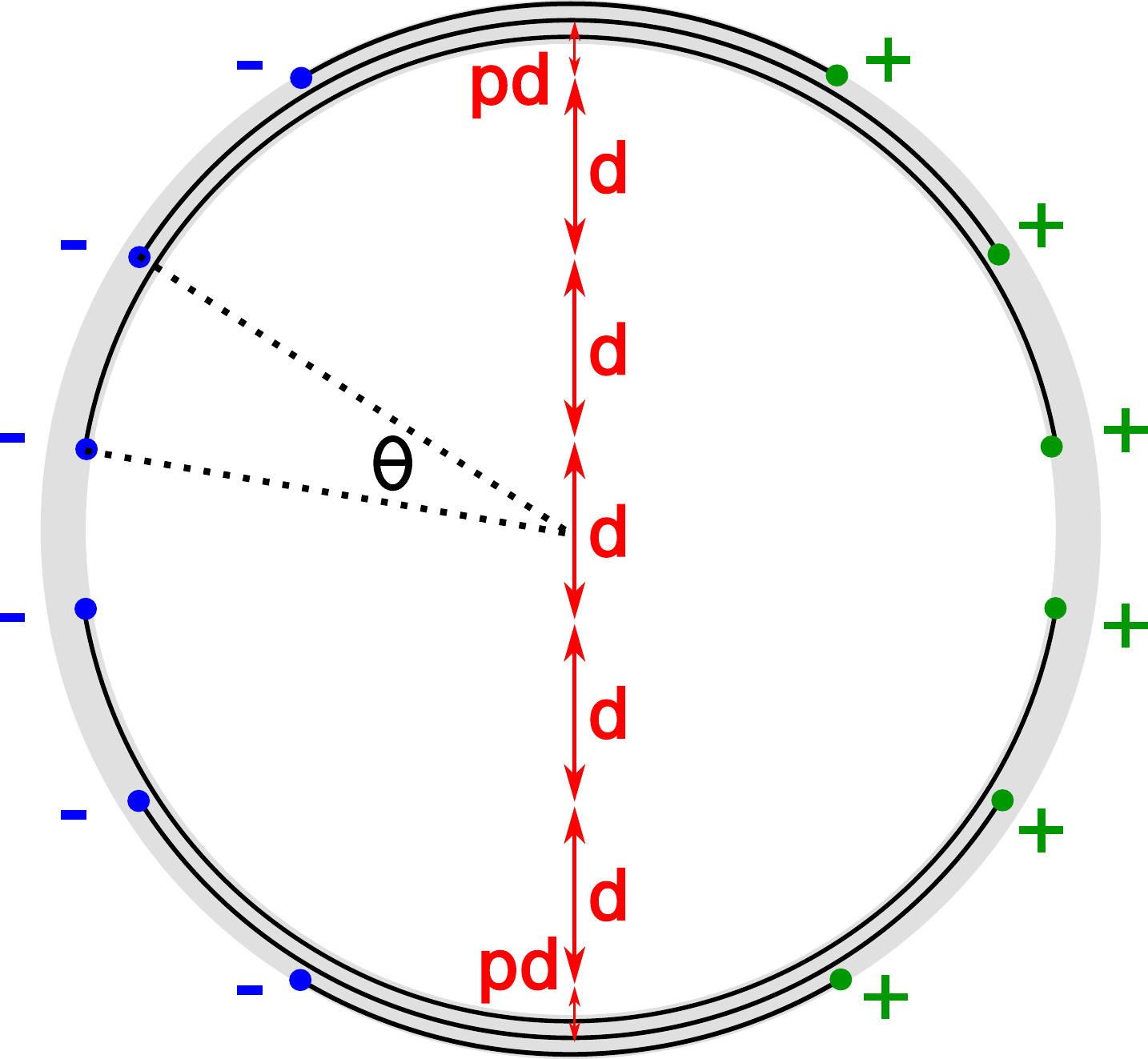}
    \caption{Visualization of the EMCS connection scheme at the circular end rings with 6 wire loops.
        The implemented $\cos\theta$ current distribution results in a homogeneous magnetic field inside the spectrometer.
        The loop distance $d$ and end parameter $p$ were optimized via simulations~\cite{Glueck2013}.
    }
    \label{fig:EMCS_current}
\end{figure}

The EMCS layout follows the `spherical cosine coil' design described in~\cite{Clark1938,Everett1966}.
In this approach, a $\cos\theta$ current distribution on the surface of a sphere is employed to create a homogeneous field in the inner volume, where $\theta$ denotes the azimuthal angle (fig.~\ref{fig:EMCS_current}).
This scheme was expanded to ellipsoidal geometries in~\cite{Everett1966,Smythe}.
Since the spectrometer vessel includes three large pump ports at the detector side of the main spectrometer (see fig.~\ref{FigMainspec}), it does not allow to implement an ellipsoidal geometry.
Instead, the EMCS was designed with a cylindrical geometry~\cite{Osipowicz2005,Glueck2013}.
The length of the EMCS cylinder is \SI{23.2}{m} and its diameter is \SI{12.6}{m}, allowing the EMCS to be mounted on the same holding structure as the LFCS (fig.~\ref{FigMainspec}).

\begin{figure}[tb]
    \centering
    \includegraphics[height=5.5cm]{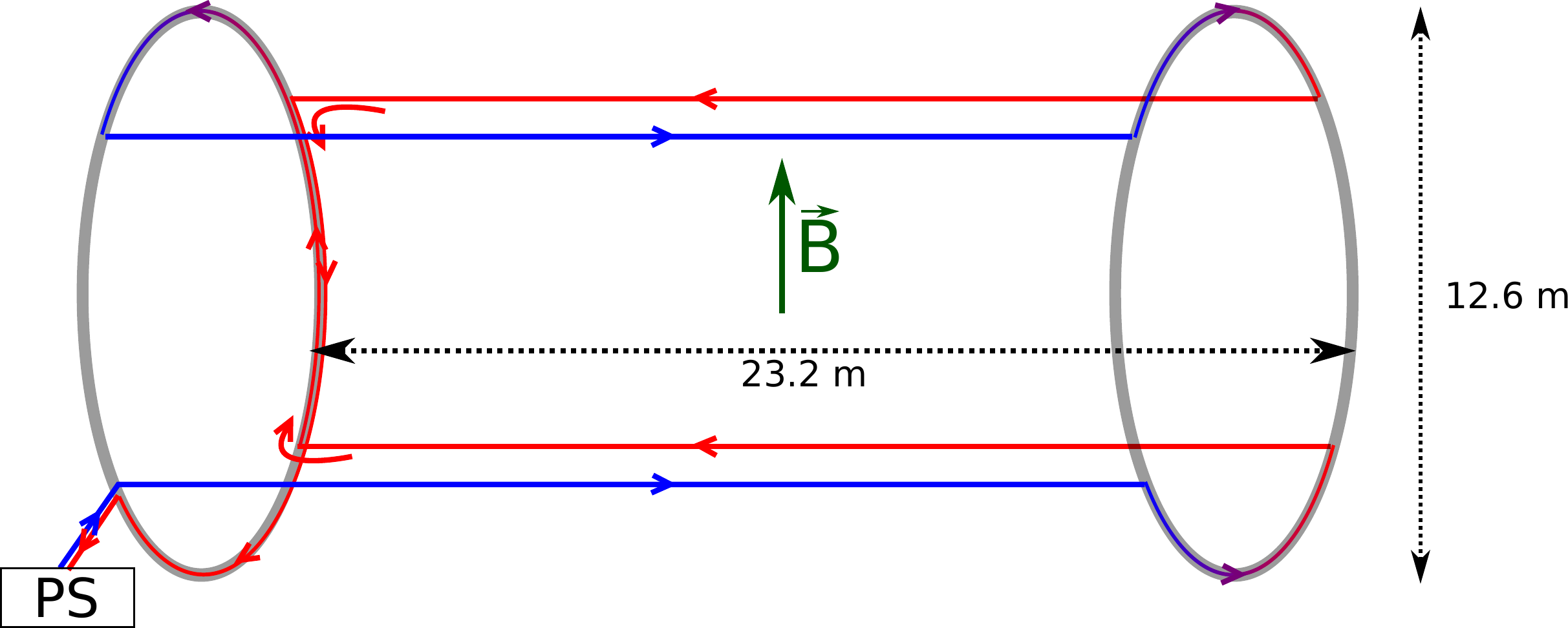}
    \caption{Visualization of the EMCS current scheme with 2 loops.
        Two horizontal current loops are united into one coil system, fed by a single power supply (PS).
        The colors and arrows indicate the direction of the current flow.
    }
    \label{fig:EMCS_total}
\end{figure}

The EMCS consists of two independent coil systems with adjustable currents that consist of several layers of wire loops along the horizontal or vertical axis (see fig.~\ref{fig:EMCS_total} for two horizontal loops).
Each loop consists of two linear current segments parallel to the spectrometer axis, which are connected at the endrings (see fig.~\ref{fig:EMCS_current} and fig.~\ref{fig:EMCS_total}).
The vertical and horizontal components (perpendicular to the beam axis) of the earth magnetic field in the KATRIN reference frame\footnote{%
    The components of the earth magnetic field are \SI{20562(138)}{nT} in north direction, \SI{788(89)}{nT} in east direction, and \SI{43827(165)}{nT} in vertical direction.
    The field has a declination of \SI{2.20(36)}{\degree} and an inclination of \SI{64.85(22)}{\degree}.
    These values are given for the location of the main spectrometer with reference date of January 1st, 2017~\cite{WMM2017}.
    The spectrometer axis itself is oriented \ang{15.4} relative to geographical north.}
amount to \SI{43.72}{\micro T} and \SI{4.90}{\micro T}, respectively~\cite{PhDErhard2016}.
The vertical component of the earth magnetic field is compensated by 16 horizontal loops of the vertical EMCS that are operated at \SI{50.0}{A}.
The smaller horizontal component is matched by 10 vertical loops of the horizontal EMCS at \SI{9.0}{A}.

\subsection{Mechanical Layout}
\label{sec:mechanical}

The current-carrying cables of the LFCS and EMCS coils are supported by a mechanical structure which has to match the following requirements:
\begin{itemize}
    \item A high geometrical precision is required to position the cables as defined by the electromagnetic design within the allowed tolerances~\cite{Glueck2013}.

    \item Non-magnetic materials have to be used to avoid magnetic field disturbances inside the main spectrometer.

    \item The structure has to carry the weight of the cables and other attachments, in addition to its own weight.
        Deformations caused by the weight must be kept within the tolerances allowed by the electromagnetic design.

    \item During measurements, a high voltage up to \SI{-35}{kV} is applied to the spectrometer vessel.
        For safety reasons, all metal parts of the grounded air coil system must keep a minimal distance of \SI{0.5}{m} from any voltage-carrying parts.
\end{itemize}

Consequently, aluminum was chosen as material for the support structure, which combines optimal mechanical (density and stiffness) and magnetic (paramagnetic) properties.
The main elements of the support structure are formed by 25 axially aligned rings, shown in fig.~\ref{FigMainspec} and \ref{fig:LFCS}.
They are labeled $\text{R1 -- R25}$, with $\text{R1}$ being located closest to the source and $\text{R25}$ closest to the detector.
Several ring structures are not used for the LFCS, but are required for mechanical stiffness and as support structure for the EMCS.
Table~\ref{TabLFCS} provides the relevant parameters of the LFCS; the nominal coil currents are given for the standard LFCS/EMCS setting which produces a field of \SI{0.36}{mT} in the center of the main spectrometer~\cite{PhDErhard2016}.
The LFCS coils $\text{L1 -- L14}$ correspond to the mechanical rings in the following way: $\text{L1}\leftrightarrow\text{R5}$, $\text{L2 -- L13}\leftrightarrow\text{R7 -- R18}$, $\text{L14}\leftrightarrow\text{R20 + R21}$; where $\text{L14}$ is constructed as a double-coil (see below).
The lower half of rings $\text{R3 + R4}$ and $\text{R23 + R24}$ are omitted, as they would interfere with the support pillars of the spectrometer hall (fig.~\ref{fig:LFCS}).

\begin{table}
    \begin{center}
        \vspace{2mm}
        \caption{
            LFCS and EMCS coil parameters:
            For the LFCS coils ($\text{L1--L14}$), $z_c$ is the axial distance of the coil center to the center of the spectrometer.
            The values $N_{\rm turns}$, $I_{\rm max}$ and $I_{\rm nom}$ denote the number of turns, the maximum current and the nominal current of the particular coil.
            $\text{L14}$ is implemented as a double-coil with two adjacent layers; it thus has two axial positions with 14 turns per sub-coil.
            All LFCS coils have an outer radius of \SI{6.3}{m} and an axial length of \SI{19}{cm}.
        }
        \begin{tabular}{lrrrrr}
            \toprule
            Coil index & $z_c$ (\si{m}) & $N_{\rm turns}$ & $I_{\rm max}$ (\si{A}) & $I_{\rm nom}$ (\si{A}) & Power supply \\
            \midrule
            L1  & \num{-6.8}           &   14              &   100 &  21.1 & SM30-100D             \\
            L2  & \num{-4.9}           &   14              &   100 &  25.7 & SM30-100D             \\
            L3  & \num{-4.0}           &   8               &   175 &  20.3 & SM30-200D             \\
            L4  & \num{-3.1}           &   8               &   175 &  28.4 & $2\,\times$ SM15-200D \\
            L5  & \num{-2.2}           &   8               &   175 &  38.8 & SM30-200D             \\
            L6  & \num{-1.3}           &   8               &   175 &  27.5 & $2\,\times$ SM15-200D \\
            L7  & \num{-0.4}           &   8               &   175 &  34.4 & SM30-200D             \\
            L8  & \num{ 0.5}           &   8               &   175 &  50.7 & $2\,\times$ SM15-200D \\
            L9  & \num{ 1.4}           &   8               &   175 &  10.4 & SM30-200D             \\
            L10 & \num{ 2.3}           &   8               &   175 &  44.4 & $2\,\times$ SM15-200D \\
            L11 & \num{ 3.2}           &   8               &   175 &  37.2 & SM30-200D             \\
            L12 & \num{ 4.1}           &   14              &   100 &  21.0 & SM30-100D             \\
            L13 & \num{ 5.0}           &   14              &   100 &  43.3 & SM30-100D             \\
            L14 & \num{ 6.6}/\num{ 6.9}&   $2\,\times$ 14  &    70 &  50.4 & SM45-70D              \\
            \midrule
            EMCS-X & ---               &   $2\,\times$ 5   &   200 &   7.2 & SM30-100D             \\
            EMCS-Y & ---               &   $2\,\times$ 8   &    70 &  46.4 & SM45-70D              \\
            \bottomrule
        \end{tabular}
        \label{TabLFCS}\label{TabEMCS}
    \end{center}
\end{table}

\begin{figure}[tb]
    \centering
    \includegraphics[height=6cm]{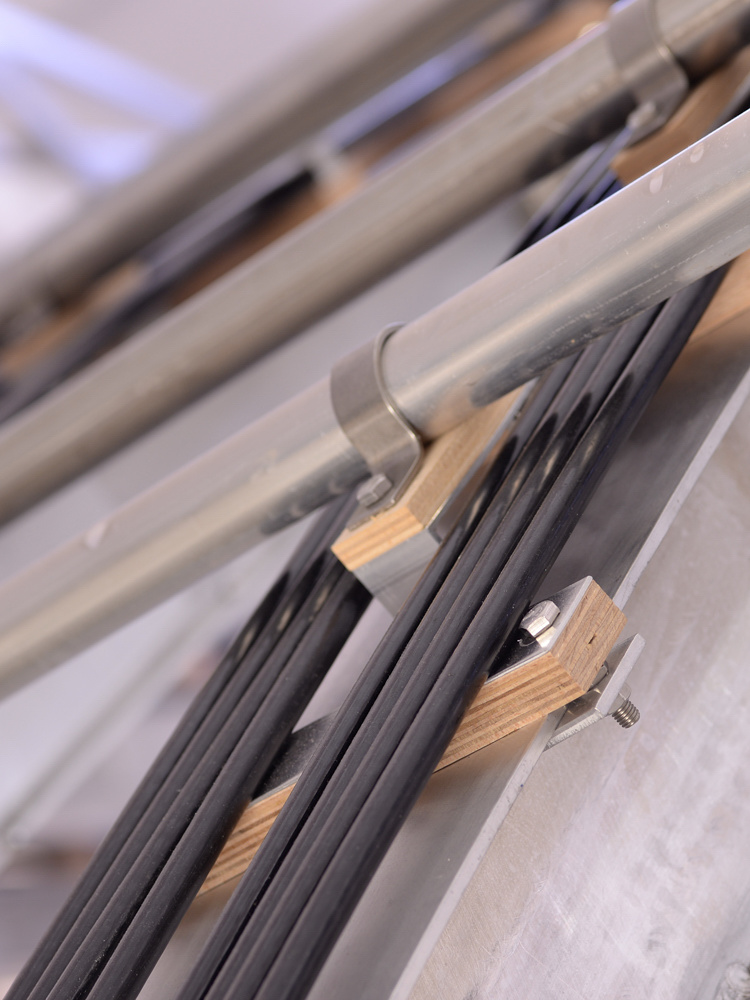}
    \quad
    \includegraphics[height=6cm]{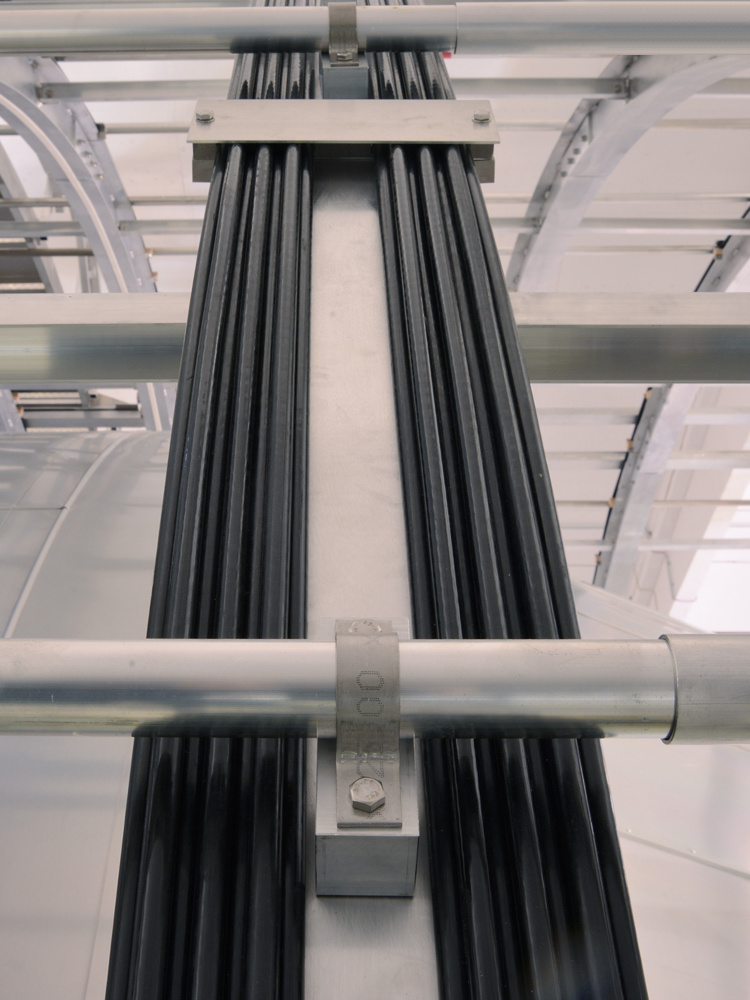}
    \quad
    \includegraphics[height=6cm]{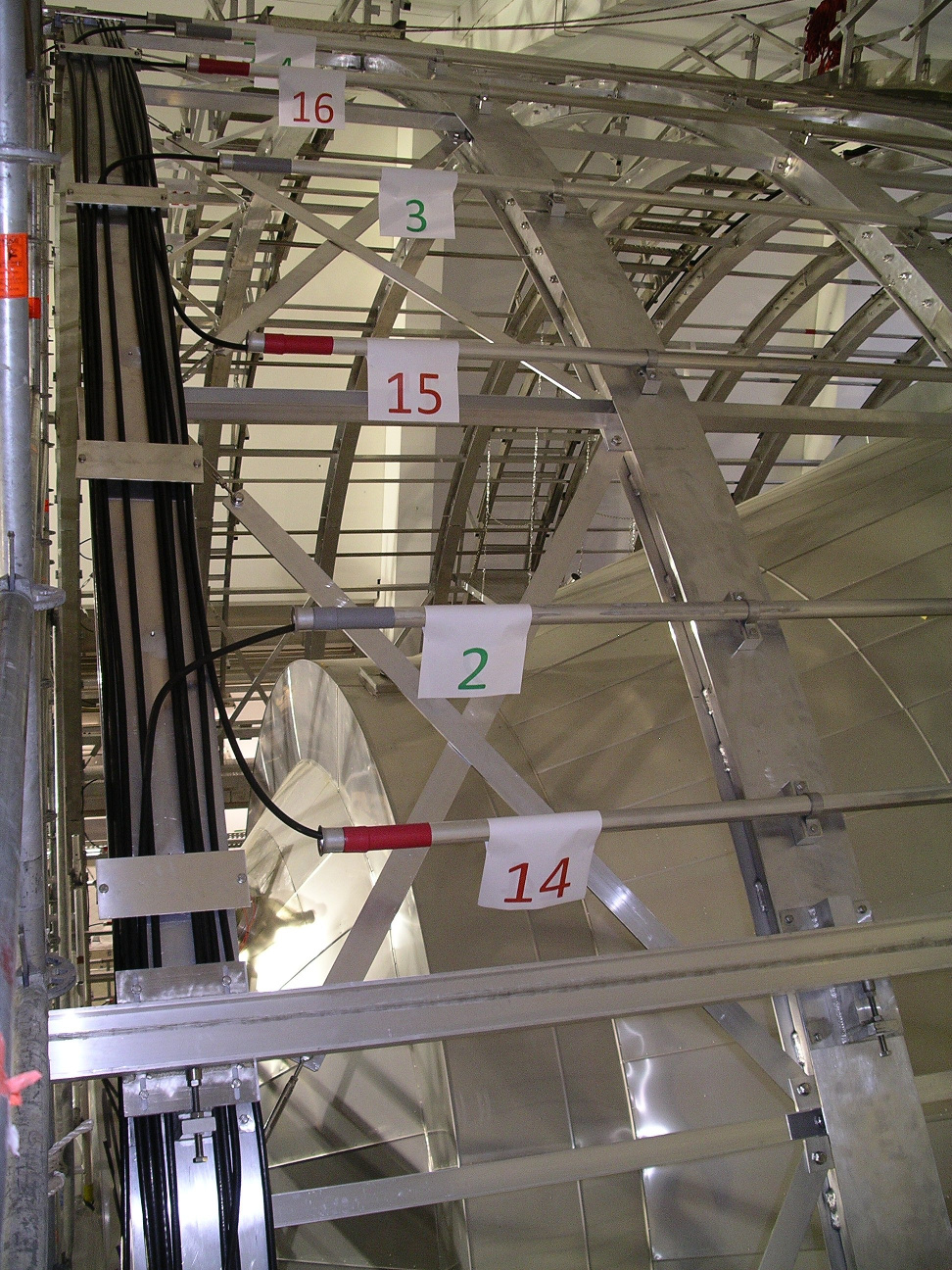}
    \caption{LFCS and EMCS cabling details.
        Left: 8-turn LFCS coil with wooden spacer and EMCS pipes running across.
        Middle: A 14-turn coil guiding 8 lower cables and 6 cables on top.
        Right: EMCS loops are completed on the end rings.
        The vertical correction (horizontal loops) is marked by red numbers, the horizontal correction (vertical loops) by green numbers.}
    \label{FigCables}
\end{figure}

Each ring consists of twelve segments made from rectangular aluminum profiles that were bent to an outer radius of \SI{6300}{mm} and an inner radius of \SI{6160}{mm}.
The profile's cross section is $\SI{80}{mm} \times \SI{140}{mm}$ with a wall thickness of \SI{4}{mm}.
Two \SI{5}{mm} thick aluminum sheets are welded onto each ring in order to form an inner and outer belt.
The \SI{205}{mm} wide outer belt carries combs with eight grooves guiding up to eight cables in a first layer (see fig.~\ref{FigCables}, left).
In case of the 14-turn coils (see fig.~\ref{FigCables}, middle, and tab.~\ref{TabLFCS}), a second layer of six cables is placed on top of the first layer.
The inner belt has a width of \SI{140}{mm} and is used to guide mobile magnetic field sensor (MobS) units~\cite{Osipowicz2012}, which are detailed below.

\begin{figure}[tb]
    \centering
    \includegraphics[height=5.5cm]{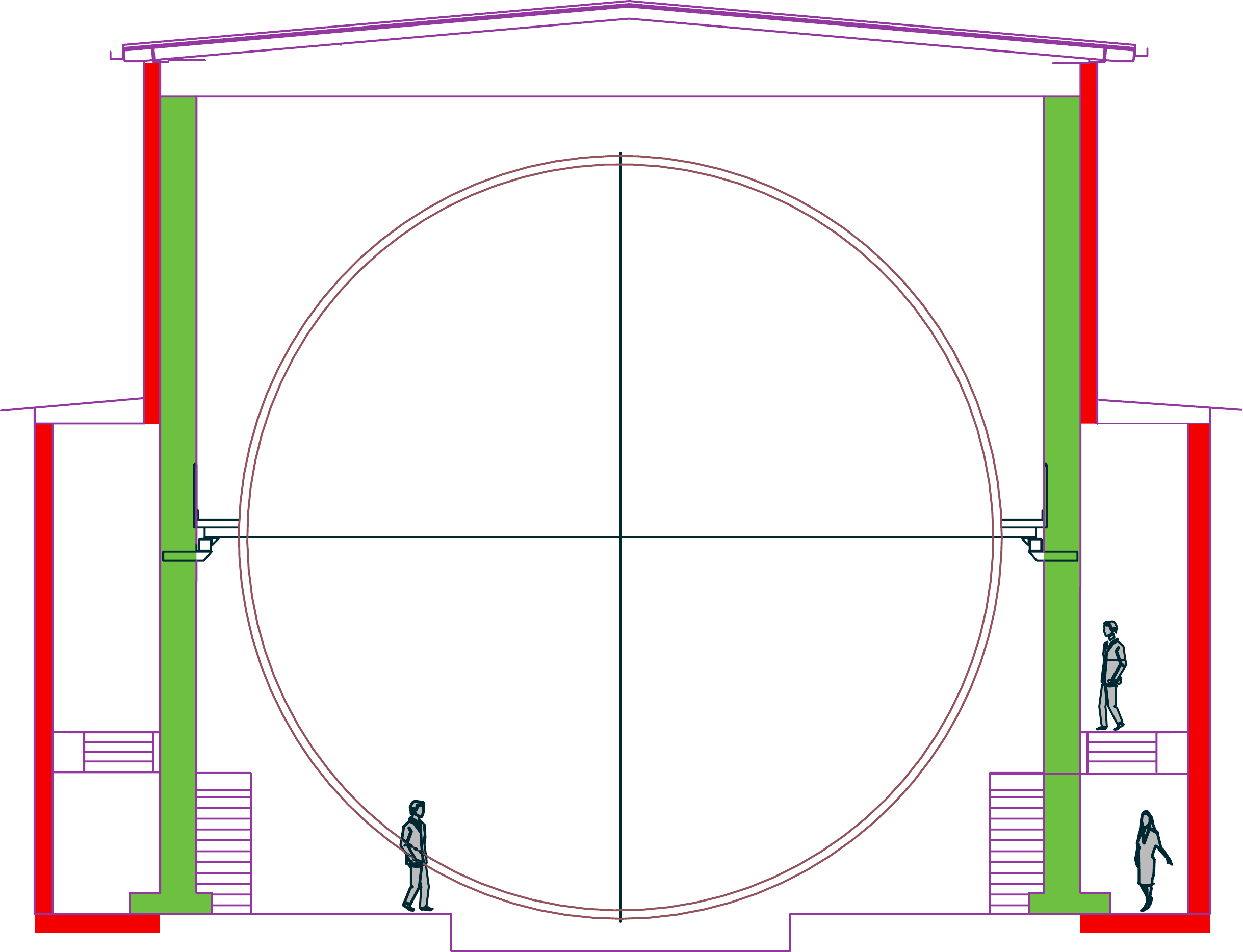}
    \quad
    \includegraphics[height=5.5cm]{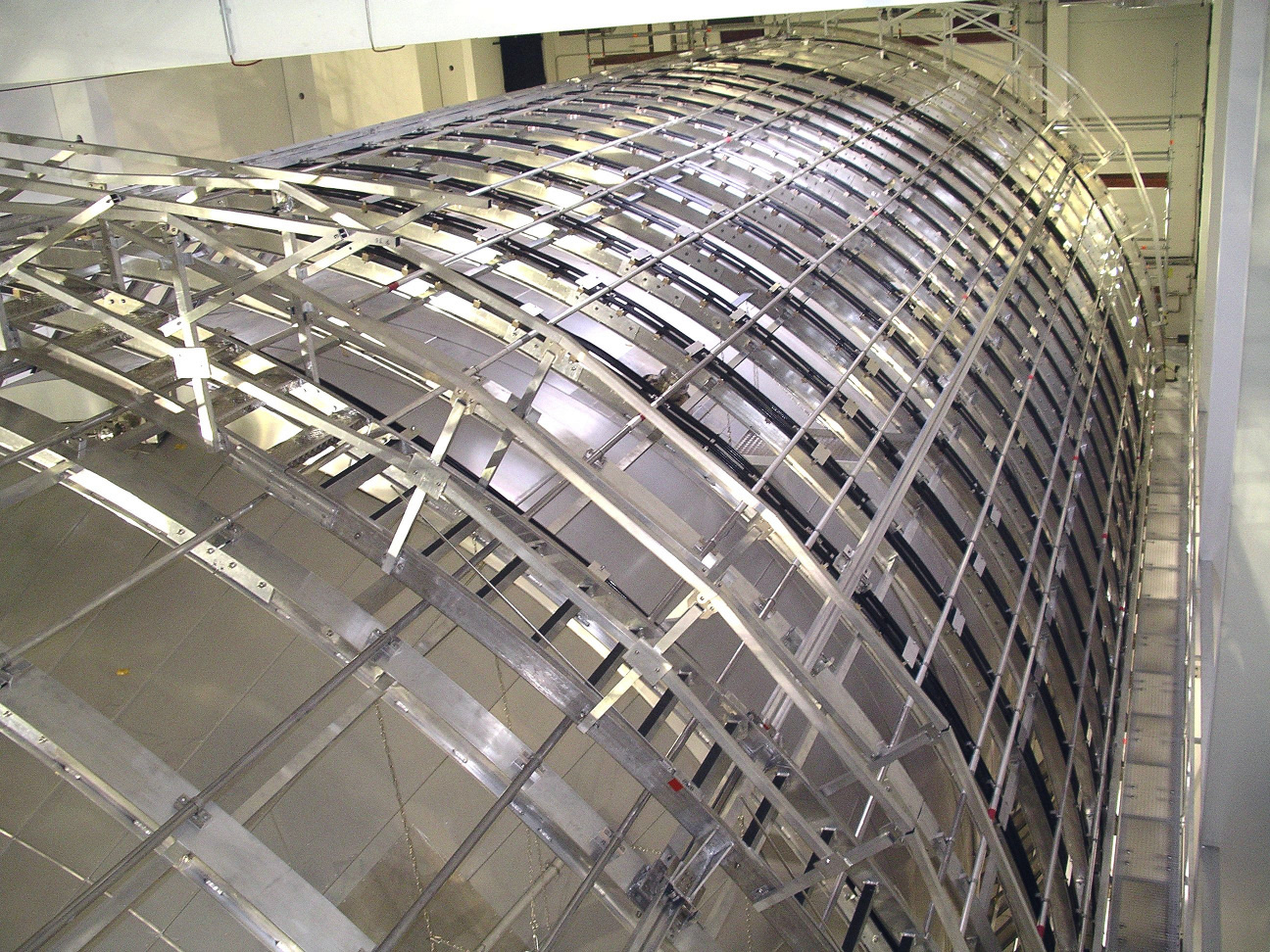}
    \caption{KATRIN main spectrometer hall and the air coil system.
        Left: The drawing illustrates how the ring system is attached to the pillars (green) of the spectrometer hall.
        The ferromagnetic reinforcement in the concrete walls is indicated in red.
        Right: Top view of the air coil system with the LFCS coils covering the central part of the main spectrometer.
        A catwalk with curved movable ladders on each side provides access to the LFCS and the spectrometer vessel.}
    \label{FigBuilding}
\end{figure}

Fig.~\ref{FigBuilding} gives an overview of the support structure of the air coil systems in the spectrometer building.
A stainless steel rail mounted at both sides of the air coil support structure at half its height transfers the weight of the entire system into the baseplate of the building.
This is achieved via the concrete support pillars of the spectrometer wall (fig.~\ref{FigBuilding}, right) that are reinforced with stainless steel.
This design avoids bulky support elements underneath the air coil rings, while also providing good accessibility to the spectrometer vessel for maintenance work.
The baseplate below the spectrometer is made of stainless steel as well.
Two catwalks are implemented on top of the air coil support structure for access to the upper half of the air coil system and the main spectrometer vessel.
Two curved aluminum ladders on both sides of the vessel can be moved along the entire length of the support structure; temporary scaffoldings are used for access to the lower half.

For the installation of the LFCS, the support rails that are attached to the pillars were extended to the detector platform using a scaffolding.
On the platform the upper half rings of the support structure were set up and moved along the rails to their intended location, facilitated by a teflon sliding surface on the rails.
After all the upper half rings reached their final position, the system was completed by installing the lower half rings with help of conventional scaffolding.

The extension of the ring system outside the LFCS is required for mechanical support of the EMCS, which is attached to the rings.
The EMCS should ideally consist of two perpendicular cosine coil systems.
However, the actual EMCS differs considerably from this ideal geometry at the endrings, where arc segments are implemented (see fig.~\ref{fig:EMCS_total}).
The end region was thus moved further out from the central low-field section of the spectrometer to reduce magnetic field distortions.

The horizontal loops of the vertical EMCS feature an equidistant vertical spacing of $\Delta y = d = \SI{0.78}{m}$, as shown in fig.~\ref{fig:EMCS_current}.
The optimal offset at the bottom and top ends of the coil system was determined via simulations~\cite{Glueck2013} to $p d = 0.6 \cdot \Delta y = \SI{0.47}{m}$.
Each loop consists of a single cable running through an aluminum pipe with a diameter of \SI{40}{mm} that is mounted parallel to the beam axis along opposite sides of the LFCS structure; the pipes are thus mirrored in horizontal direction.
The inner diameter of these pipes constrains possible deviations of the cable from its reference position to $< \SI{11}{mm}$, which suffices for the earth magnetic field compensation.
At the end rings $\text{R1 + R25}$ the cables leave the pipes and run along the circular outside of the rings, similar to the LFCS cables, either closing the loop or proceeding to the next loop (see fig.~\ref{FigCables}, right).
The end ring at the detector side (north, $\text{R25}$) closes the loops by connecting segments on opposite sides of the vessel, while the source-side ring (south, $\text{R1}$) carries the supply leads and connects adjacent loops of each coil system.
Consequently, the interconnected loops form an entire cosine coil for each component (horizontal and vertical) of the EMCS.

The vertical loops of the horizontal EMCS have a distance of $\Delta x = \SI{1.24}{m}$ in horizontal direction and an offset of $0.6 \cdot \Delta x = \SI{0.74}{m}$ to the sides of the ring system.
One pipe of each loop is located in the bottom half of the support structure and the corresponding partner pipe is located vertically above in the top half; the pipes are thus mirrored in vertical direction.
Here only ten loops are required because the horizontal component of the transversal earth field is much weaker than the vertical one.
The remaining earth magnetic field component in direction of the spectrometer axis (beam axis) is accounted for by the LFCS~\cite{PhDErhard2016}.
As shown in fig.~\ref{FigCables}, the loops of the vertical EMCS carry red numbers \numrange{1}{16} and those of the horizontal EMCS have green numbers \numrange{1}{10} for easy identification.

\subsection{Electrical layout}
\label{sec:electrical}

There are two options for the conductor material of the air coil cables: copper and aluminum.
A comparison of cable geometries with an identical resistance per unit length of about \SI{0.4}{\Ohm/m} shows that aluminum offers the advantage of a considerably lighter and cheaper material (see tab.~\ref{TabAlCu}).
A copper cable of same resistance would be more than twice as heavy and about three times as expensive\footnote{%
    A \SI{70}{mm^2} aluminum cable has a resistance of \SI{0.403}{\Ohm/m} and a mass of \SI{189}{g/m}.
    A copper cable with same resistance would have a smaller cross-section of only \SI{42}{mm^2}, but still be considerably heavier at \SI{375}{g/m}.}.
The difference in weight is of particular importance for the statics of the air coil system, with copper requiring a stronger mechanical support structure.
As this structure is by far the most expensive part of the air coil system, aluminum was selected as conductor material for both air coil systems.

The aluminum cables consist of an inner 19-wire conductor with a total cross-section of \SI{70}{mm^2} and an outer polyethylene (PE) insulation layer of \SI{1.5}{mm} thickness (model BayEnergy NFA2X~1$\times$70~RM~0.6/1kV~\cite{BayEnergy2013}).
This type of cable is widely used for outdoor above-ground power distribution and reliable connection methods are easily available, which is of particular concern with aluminum conductors.
The cable also offers the necessary stiffness for a stable mechanical design, while still being flexible enough to form wire loops and air coils.
Given the mechanical design, the effective radius of a conductor cable on a LFCS ring is \SI{6.317}{m}.
The maximum current for the outdoor cables running free between posts is \SI{205}{A}; the LFCS and EMCS are operated at lower currents up to \SI{175}{A}~\cite{PhDErhard2016,PhDBehrens2016}.

The resistance $R_N$ of a coil with $N$ turns, a conductor cross section $A = \SI{70}{mm^2}$ and a coil diameter $D = \SI{12.6}{m}$ at a temperature of \SI{20}{\celsius} can be calculated from the resistivity given in tab.~\ref{TabAlCu} and the adapted standard equation
\begin{equation}
    \label{eq:coil_resistance}
    R_N = \rho \cdot \frac{N \cdot \pi D + L_\mathrm{S}}{A}
    \,,
\end{equation}
where $L_\mathrm{S} \approx \SI{20}{m}$ denotes the length of the supply lead from the coil to the power supply unit.
For the coils with $N = \num{8}$ turns we obtain a theoretical value $R_8 = \SI{0.136}{\Ohm}$ for a cable length of \SI{317}{m} for the coil windings.
The values actually measured for the 8-turn LFCS coils at \SI{20}{\celsius} vary between \SI{0.1429}{\Ohm} and \SI{0.1458}{\Ohm} with an average of \SI{0.1439}{\Ohm}.
The variance observed in the measurements results to a large extent from the varying lengths of the supply leads for individual coils.
The difference between measurements and calculation stems from the fact that \eqref{eq:coil_resistance} does not take into account the internal 19-wire structure of the conductor, which slightly increases the resistance.
The measured average resistance for the 8-turn coil corresponds to \SI{0.427}{\Ohm/km}, which is in agreement with the maximum value of \SI{0.443}{\Ohm/km} quoted by the manufacturer~\cite{BayEnergy2013}.

Each coil is individually regulated by power supplies of the SM-3000 series by manufacturer Delta Elektronika BV~\cite{DeltaSM3000} (tab.~\ref{TabLFCS}).
This series includes power supplies with different combinations of maximum available current and voltage, resulting in a maximum power output up to \SI{6}{kW}.
The model SM30-100D of this series delivers up to $I_\mathrm{max} = \SI{100}{A}$ at a maximum voltage of $U_\mathrm{max} = \SI{30}{V}$.
Due to the internal resistance of the air coils, the maximum current is also limited by the voltage drop in the load circuit.

For the 8-turn coils $\text{L3--L11}$, currents up to \SI{175}{A} are possible.
This is achieved by using either one SM30-200D power supply or two SM15-200D power supplies in series.
Both options deliver a current up to \SI{200}{A} at $U_\mathrm{max} = \SI{30}{V}$ and are used at the LFCS.
For the second option, a master-slave series adapter by Delta Elektronika is used for proper power distribution.
In a test measurement one of the central LFCS coils was operated at maximum current ($\approx \SI{180}{A}$) for several days at an ambient hall temperature of \SI{18}{\celsius}.
During this test, the surface temperature of the cable did not exceed a maximum of \SI{60}{\celsius} with typical temperatures around \SI{40}{\celsius} over a large coil section.
These temperatures are well within the specificied limits and no significant increase in resistance was observed.

For the 14-turn LFCS coils $\text{L1+L2}$ and $\text{L12+L13}$ we also use the SM30-100D model. Here the coil temperature increases by up to \SI{20}{\celsius} at \SI{100}{A}.
With a measured resistance of \SI{0.247}{\Ohm} at \SI{20}{\celsius}, the required voltage for a maximum current of \SI{100}{A} is \SI{24.7}{V}; the power supply with $U_\mathrm{max} = \SI{30}{V}$ provides ample reserve to account for the resistance increase by temperature.
The same margin applies to the 28-turn LFCS coil $\text{L14}$ with a resistance of \SI{0.507}{\Ohm}; this double-coil is formed by two adjacent 14-turn coils that share the same power supply (see tab.~\ref{TabLFCS}).
Here a SM45-70D model with $U_\mathrm{max} = \SI{45}{V}$ is used to deliver the required voltage of \SI{35.5}{V} at \SI{20}{\celsius} for the maximum current of \SI{70}{A}.

The EMCS uses the same type of aluminum cable as the LFCS.
In standard operation mode, the 16-turn coil $\text{E-Y}$ for the vertical field compensation operates at \SI{46.5}{A}, while the 10-turn coil $\text{E-X}$ for the horizontal field compensation operates at \SI{7.1}{A}; this setting incorporates the influence from magnetic materials and are thus different from the design values~\cite{PhDErhard2016}.
For test measurements and to correct the position of the flux tube inside the spectrometer vessel, the EMCS must be able to shift the flux tube in horizontal and vertical direction by up to \SI{0.5}{m} in the analyzing plane.
This is achieved by changing the current by up to \SI{25}{A} (\SI{44}{A}) for the horizontal (vertical) coil, resulting in a maximum current of \SI{75}{A} and \SI{53}{A}, respectively\footnote{%
    For the horizontal coil, a shift in negative direction requires a current of \SI{-35}{A}.
    The negative sign is achieved by inverting the coil polarity with the flip-box system (sec.~\ref{sec:magpulse}).}.
In case of the vertical EMCS with a SM45-70D power supply, the \SI{75}{A} requirement is only marginally limited by the power supply with $I_\mathrm{max} = \SI{70}{A}$.
The resistance of the 16-turn coil was measured to \SI{0.508}{\Ohm}, resulting in a maximum voltage of \SI{35.6}{V} at a current of \SI{70}{A} that is deliverd by a SM45-200D power supply.
The 10-turn coil with a resistance of \SI{0.311}{\Ohm} is operated by a SM30-100D power supply, which delivers the required voltage of \SI{16.5}{V} at a current of \SI{53}{A}.

With the integration of the magnetic pulse system at the main spectrometer (sec.~\ref{sec:magpulse}), the polarity of all LFCS and EMCS coils can be inverted without any hardware changes~\cite{PhDBehrens2016}.
The polarity change is achieved by 16 current-inverter units (the so-called ``flip-boxes'') that are placed between each air coil and its corresponding power supply.
Hence, the implementation of this system did not require any additional hardware upgrades.

\begin{table}[tb]
    \begin{center}
        \vspace{2mm}
        \caption{Comparison of aluminum and copper for electrical resistivity, density and price~\cite{SerwayJewett,CRC}.}
        \begin{tabular}{llrr}
            \toprule
            material                            &                           & aluminum      &      copper       \\
            \midrule
            resistivity at \SI{20}{\celsius}    & (\si{\Ohm.mm^2/m})        &     0.0282    &      0.0168       \\
            density                             & (\si{g/cm^3})             &     2.70      &      8.92         \\
            price in 2008                       & (Euro/\si{kg})            &     5.70      &      7.80         \\
            \bottomrule
        \end{tabular}
            \label{TabAlCu}
    \end{center}
\end{table}

\subsection{Slow control}
\label{sec:sslowcontrol}

The readout and control infrastructure is based on so-called cRIO (compact Remote Input-Output) field controllers by National Instruments.
The core of the KATRIN air coil control system is a cRIO bank that is based on the \textit{cRIO-9068} Linux embedded real-time controller that allows a readout interval down to \SI{200}{ms}~\cite{PhDErhard2016}.

The communication with the measurement devices is performed by several interconnected modules:
\begin{itemize}
    \item Two single-ended \textit{cRIO-9264} ($2 \times 16$~channels with 16-bit precision) voltage-output DAC units send the current set values to the power supply's analog interfaces.

    \item Three \textit{cRIO-9205} voltage-input ADC units ($3 \times 16$~channels with 16-bit input precision) are used to read out the coil currents, the output voltages of the power supplies and the temperatures of the flip-box units.
    The currents are measured via precision current transducers by LEM~\cite{LEMIT200S}.
    The modules are isolated to prevent ground loops that could lead to adverse effects, such as a correlated (and thus inaccurate) readout.

    \item Two \textit{cRIO-9403} digital input modules (32~channels), are used to read status information from all power supplies, current transducers and flip-boxes.

    \item One \textit{cRIO-9476} digital output module (16~channels) is used to set the output polarity of the flip-boxes.
\end{itemize}

Two combined multi-channel proportional integral derivative (PID)-loops regulate the output currents of the power supplies in ramping and in constant-current (CC) mode.
The internal control mode is switched `ramping' by the slow-control software when the difference between the current set-point and the readout value exceeds a margin of \SI{5}{A}; otherwise the power supply operates in CC mode.
All slow-control data is collected, processed and archived in a central SQL database to be available for analysis~\cite{PhDKleesiek2014}.

The air coil system is linked to the safety-related control infrastructure of KATRIN via an inter-lock system.
It performs a remote shutdown of all power supplies in case of any high voltage or vacuum alarm in the main spectrometer hall.

\subsection{Magnetic field monitoring and modeling}
\label{sec:monitoring}

The transmission properties of the main spectrometer for electrons with energies close to $E_0 = \SI{18.6}{keV}$ are of key importance for the neutrino mass analysis~\cite{KATRIN2005}.
Hence, a precise knowledge of the magnetic field in the main spectrometer (especially the analyzing plane) is mandatory.
During standard operation, magnetic field sensors cannot be installed directly in the main spectrometer flux tube as they would block signal electrons.
However, the interior field can be assessed by a simulation model that is based on multiple external magnetic field sensors that are distributed around the analyzing plane.
This model takes into account the individual contributions from the superconducting solenoids in the KATRIN beamline, the large-volume air coil system, the earth magnetic field and surrounding magnetic materials.

A total of 14 high-precision magnetometers manufactured by Bartington~\cite{BartingtonMag03} are mounted directly on the surface of the spectrometer vessel.
These stationary sensors continuously measure the magnetic field with a sampling rate down to \SI{200}{ms}~\cite{PhDErhard2016} and a precision well below the required 1\% requirement for a \SI{0.3}{mT} magnetic field.
To get accurate results from the field model, the magnetic materials in the spectrometer hall (fig.~\ref{FigBuilding}) need to be included (see sec.~\ref{SecPerformance}).

In order to obtain a large ensemble of field measurements close to the main spectrometer surface within the LFCS, the inner belts of the air coil support structure also serve as trackways for mobile magnetic field sensor units (MobSU)~\cite{Osipowicz2012}.
Each MobS unit contains two fluxgate sensors\footnote{
    We use a custom design of the sensor \textit{FLC3-500} by Stefan Mayer Instruments. It features an extended measuring range of $\pm \SI{1000}{\micro T}$ and a small offset error $< \SI{50}{nT}$ at \SI{18}{\celsius}.}
with an axial distance of \SI{0.45}{m} and takes $2 \times 60$ magnetic field samples along the track at \ang{6} intervals in less than \SI{10}{min} (in standard mode).
The measurement data is stored in the slow-control database when the units reach their docking station (measurement point 0) after a full turn; during the measurement the data is stored locally on each unit.

A total of four MobS units have been installed at the LFCS coils $\text{L3}$, $\text{L6}$, $\text{L9}$ and $\text{L12}$ and have been successfully commissioned.
They create the radial magnetic monitoring system (RMMS) that allows to investigate the homogeneity of the magnetic field in vicinity of the main spectrometer, which can be disturbed by the influence of magnetic materials.

In addition to the RMMS, a dedicated system to monitor magnetic materials has been developed as well.
The vertical magnetic field monitoring system (VMMS) consists of six support structures for mobile magnetic field sensor units that follow the MobSU design and use the same type of fluxgate sensors.
The units move in vertical and horizontal direction, thus creating a mesh of measurement points with a grid size of \SI{20x20}{cm} that is oriented in parallel to the spectrometer axis.
The system was recently set up next to the support pillars of the main spectrometer hall at different heights and is awaiting full commissioning.
It will record detailed two-dimensional maps of the residual magnetic field with several thousand measurement points and thus provide new insights into the influence of magnetic materials.

\subsection{Mechanical deformation and correction}
\label{sec:gravity}

\begin{figure}[tb]
    \centering
    {\includegraphics[width=0.47\textwidth]{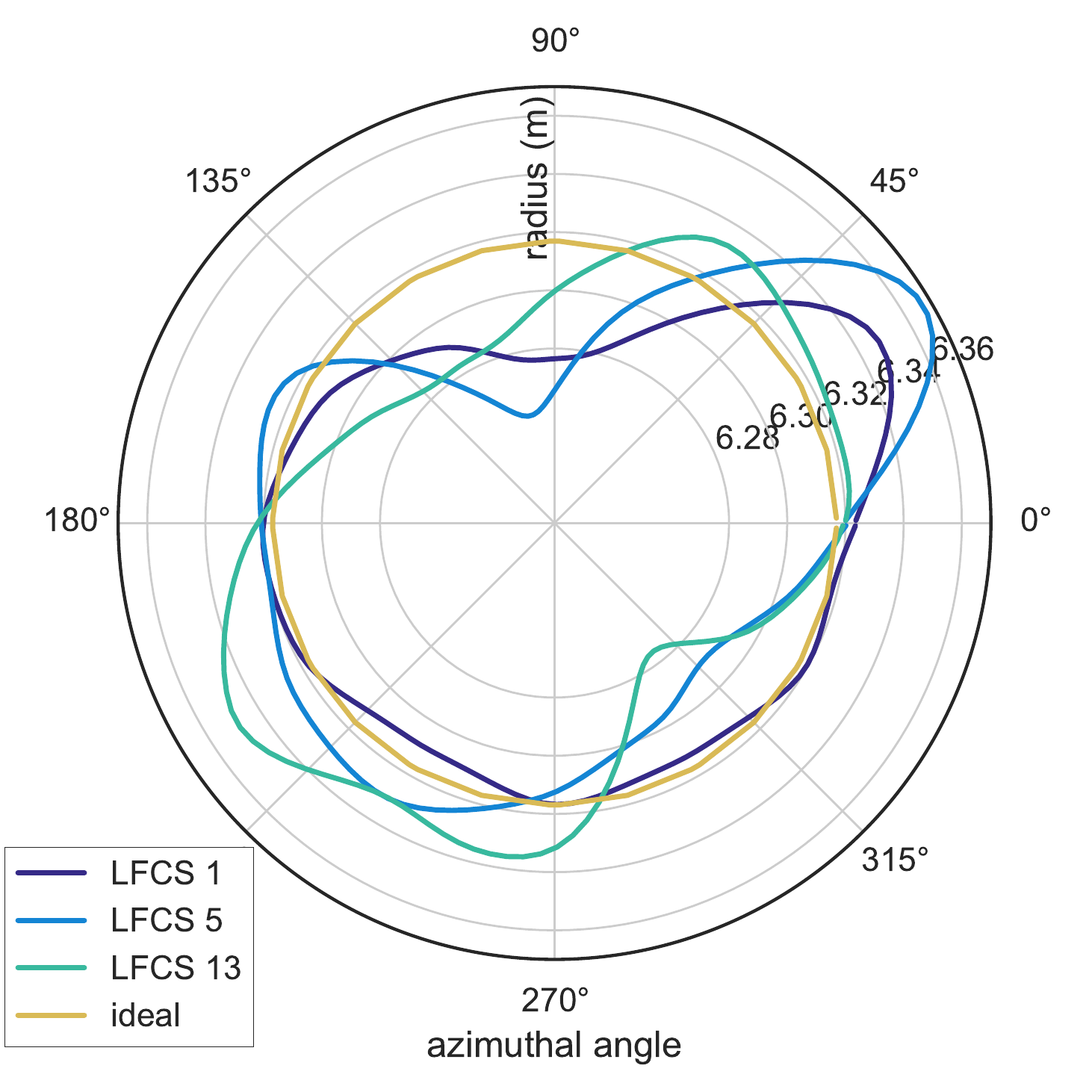}}
    \quad
    {\includegraphics[width=0.47\textwidth]{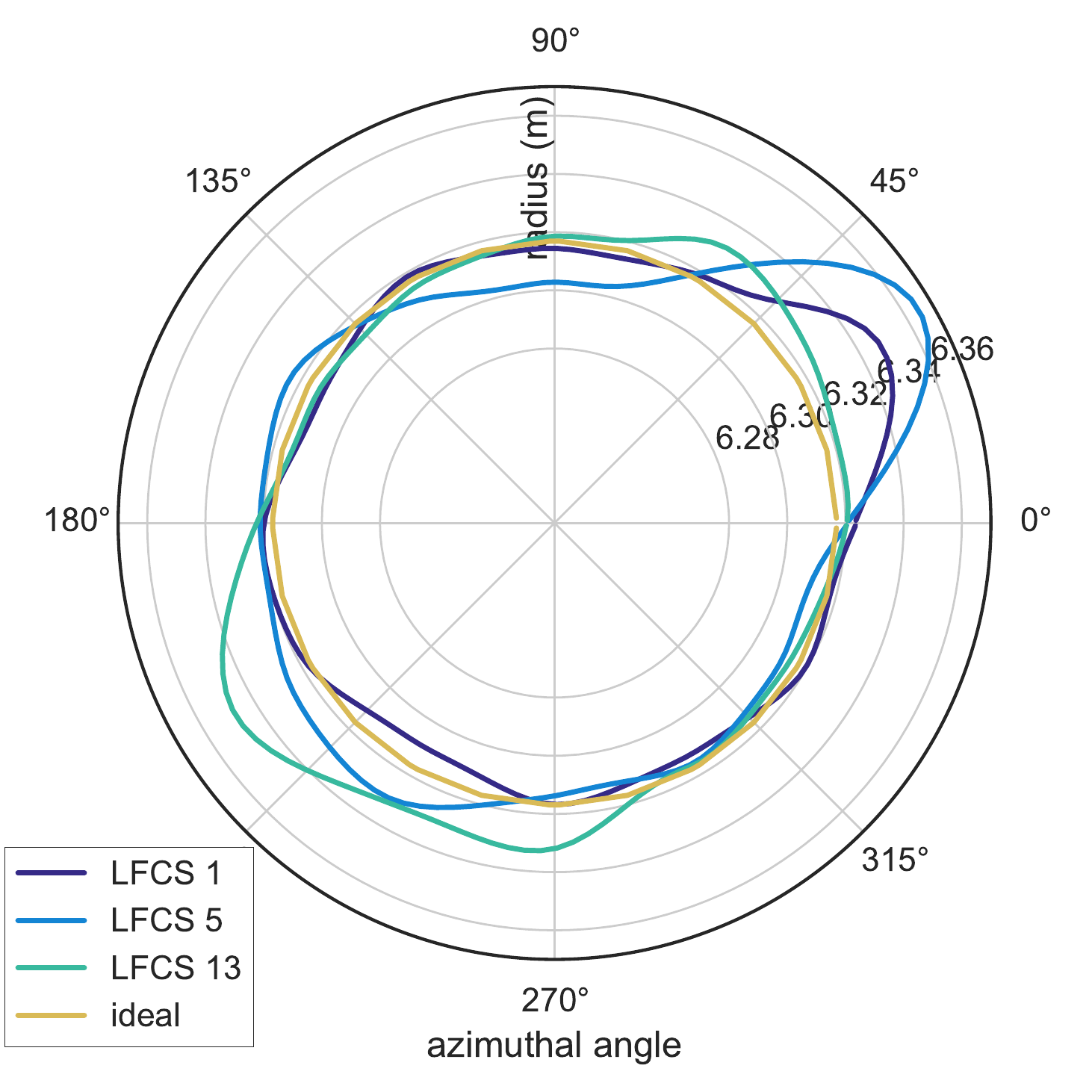}}
    \caption{Radial deviations due to mechanical deformation of LFCS coils $\text{L1}$, $\text{L5}$ and $\text{L13}$.
        The measured radius is extrapolated from the inner radius of the support structure (\SI{6155}{mm}) to a nominal outer radius of \SI{6317}{mm} for the coil windings.
        Left: A maximum deviation of \SI{50}{mm} from the expected radius is observed.
        Right: Wooden spacers were introduced to correct the wire positions by shifting the cables outwards where necessary.
        This improves any radial deviations in the inwards direction, but cannot compensate outwards deviations where the coil radius is too large.
    }
    \label{fig:grav_deform}
\end{figure}

The significant intrinsic weight and large radius of the air coil cables and their support structure result in a mechanic deformation of the coil loops.
The coils deviate from the ideal circular shape, as shown in the left panel of fig.~\ref{fig:grav_deform}.
Radial deviations from the expected radius of $r = \SI{6317}{mm}$ of up to \SI{50}{mm} are observed.
The radius was measured with a \textit{Leica TCRA 1203} laser tracker system on 36 points along LFCS rings $\text{L1}$, $\text{L5}$ and $\text{L13}$~\cite{Gumbsheimer2013}.

To bring the cables as close as possible into a circular form, wooden spacers with heights of $\SIlist{10;20;30;40}{mm}$ were installed between the cables and the support structure.
This effectively increases the coil radius and thus allows to compensate inwards deviations where the radius is lower than expected; outwards deviations cannot be compensated by this method.
After spacer installation the absolute position of the cables were determined again by a laser tracker with an accuracy of $\pm \SI{5}{mm}$ (fig.~\ref{fig:grav_deform}, right panel).
These alignment data are incorporated into the magnetic field simulations.
At some points, the reference positions of the EMCS cables collide with the four mechanical support pillars of the main spectrometer (see fig.\ref{FigMainspec}); here alternative cable routings were implemented.
The modified routings were optimized with simulations in order to minimize the resulting deviations in the analyzing plane~\cite{PhDReich2013}.

\begin{figure}[tb]
    \centering
    \includegraphics[width=1.0\textwidth]{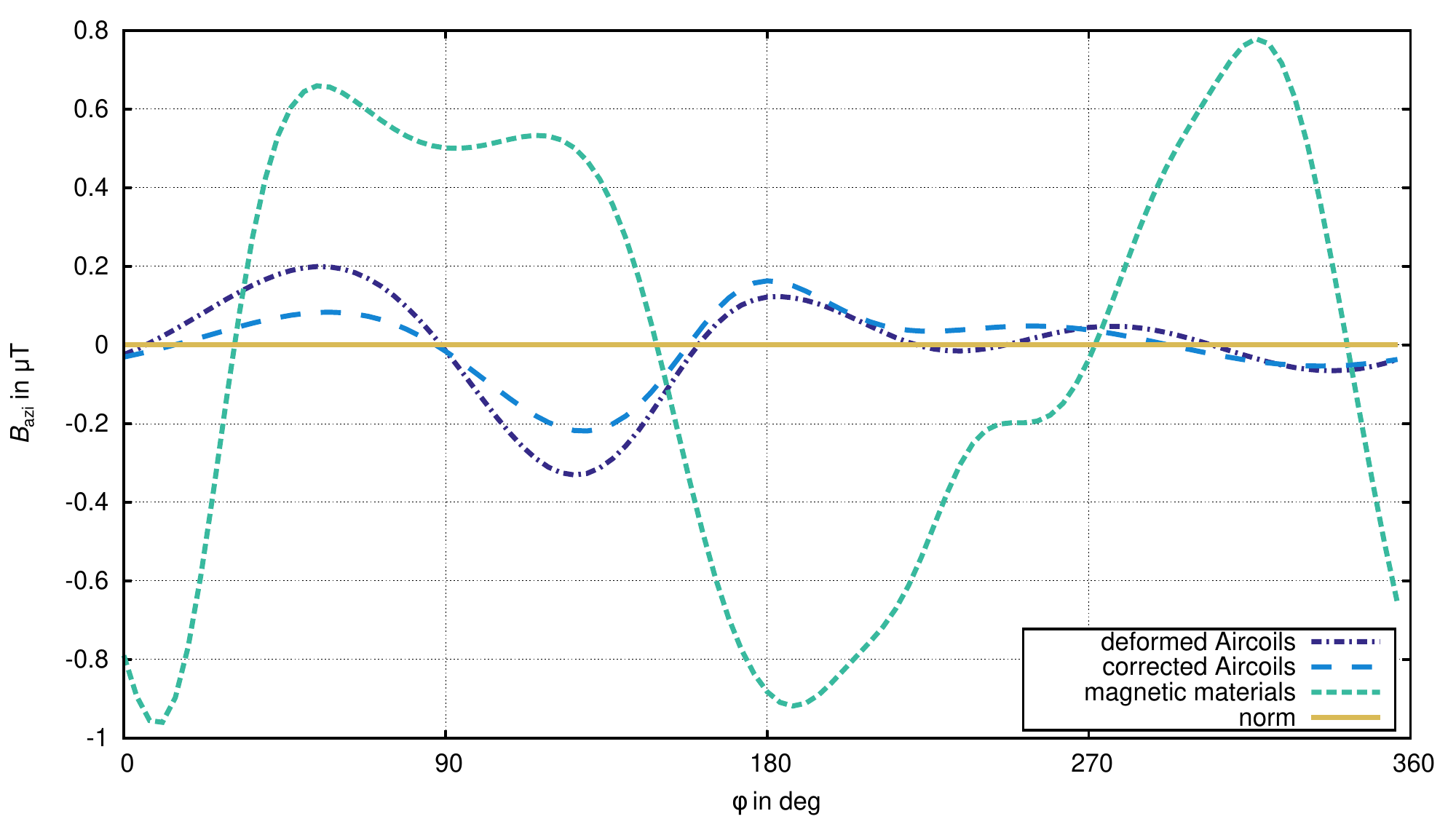}
    \caption{Azimuthal magnetic field variations at $z = 0$, $r = \SI{3.5}{m}$ from different field-disturbing sources.
        The magnetic field inhomogeneities introduced by the air coil system are an order of magnitude smaller than the disturbances from magnetic materials in the spectrometer hall.}
    \label{FigRingfield}
\end{figure}

Figure~\ref{FigRingfield} shows the influence on the magnetic field in the analyzing plane due to these deviations from the original design.
The figure includes a direct comparison to magnetic field disturbances introduced by the magnetic materials in the spectrometer hall.

\subsection{Magnetic field layout}
\label{sec:field_layout}

\begin{figure}[tbp]
    \centering
    \includegraphics[width=1.0\textwidth]{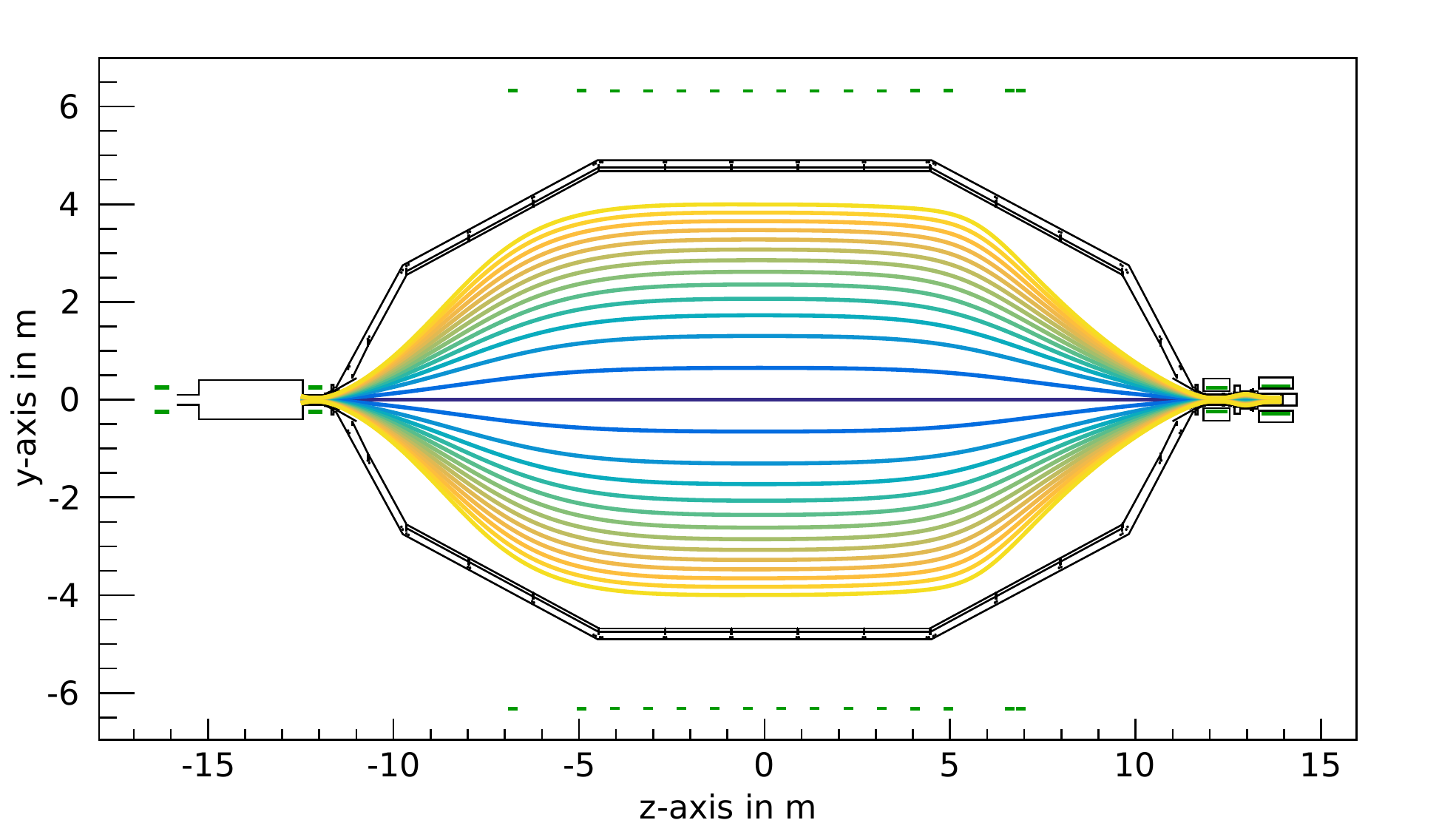}
    \\
    \includegraphics[width=1.0\textwidth]{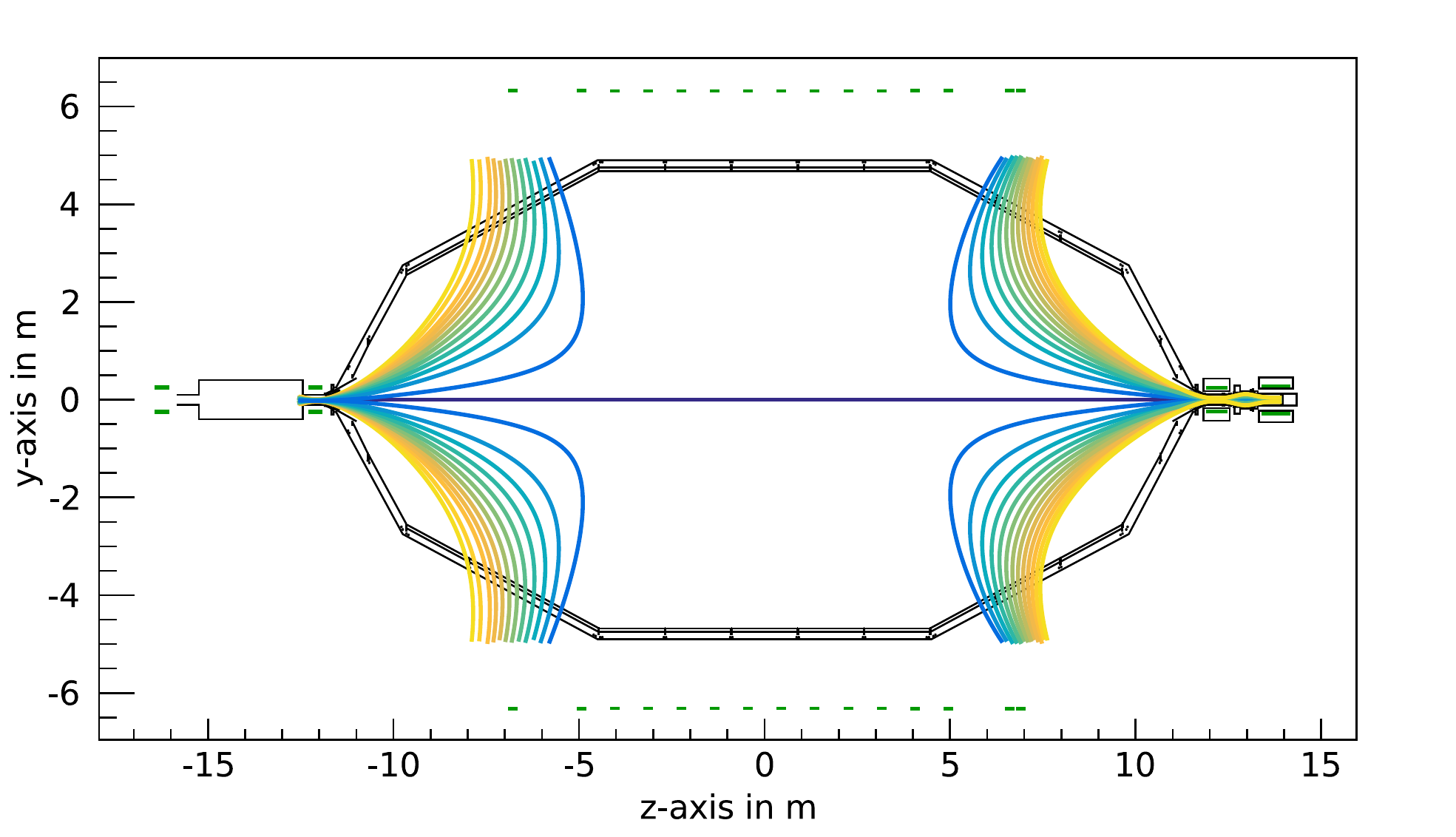}
    \caption{Simulated magnetic field in the main spectrometer.
        Top: An air coil setting close to the nominal magnetic field configuration of the MAC-E filter with about \SI{0.36}{mT} in the center of the analyzing plane.
        The depicted field lines correspond to the radial segmentation of the FPD detector.
        Bottom: The polarity of LFCS air coils $\text{L1--L13}$ are inverted by the flip-box units.
        This configuration removes stored charged particles from the spectrometer volume.
        It can also be used to investigate electron emission from the vessel walls at the detector side.}
    \label{Fig:BSettingWithPuls}
\end{figure}

The air coil system can be used to modify the magnetic field strength and even invert the field direction inside the spectrometer in a large volume around the analyzing plane.
The operation of the main spectrometer with ideal MAC-E filter characteristics is depicted in the upper panel of fig.~\ref{Fig:BSettingWithPuls}, based on an air coil current setting determined by the optimization routine described in~\cite{Glueck2013,PhDGroh2015}.
The presented case shows a magnetic field close to the nominal strength of \SI{363}{\micro T} in the center of the spectrometer~\cite{PhDErhard2016}.

During commissioning measurements of the spectrometer and detector section, the ability to adjust the magnetic field at the main spectrometer was of central importance to background investigations~\cite{PhDHarms2015}.
It is most likely that the dominant background component in the spectrometer arises from ionization processes of neutral messenger particles, which are distributed homogeneously over the entire flux tube volume (see~\cite{PhDTrost2017} for a detailed description).
A change of the flux tube volume according to \eqref{eq:fluxtube} is therefore of vital importance to reduce background; this can be achieved by applying different LFCS current settings.

A second major characteristic of the LFCS is its ability to operate the spectrometer in a mode where the electromagnetic requirements for electron transmission (sec.~\ref{sec:EMD}) are broken.
This feature of the LFCS has proven to be of central relevance for background investigations.
Here we distinguish between the inverted and the asymmetric mode:
In the \emph{inverted mode} the nominal LFCS polarities are switched, resulting in a magnetic field direction inside the spectrometer that is anti-parallel to the solenoid stray fields (lower panel of fig.~\ref{Fig:BSettingWithPuls}).
The inversion can be applied dynamically for a short time to remove stored electrons from the main spectrometer (sec.~\ref{sec:magpulse}).
A static inversion of the magnetic field allows dedicated measurements that map surface processes at the spectrometer vessel and its inner electrode system to the detector~\cite{PhDHarms2015,PhDErhard2016}.
A similar configuration can be achieved in the \emph{asymmetric mode}, where the PS1 and PS2 solenoids at the source side of the main spectrometer are turned off (not shown).
In both cases electrons from the electrode surfaces are guided to the focal-plane detector, thereby providing an alternative method for background investigations~\cite{PhDLeiber2014}.

\subsection{Active background reduction}
\label{sec:magpulse}

The inverted mode allows to invert the magnetic field inside the spectrometer within a short time ($< \SI{1}{s}$) as an active countermeasure against background from stored electrons.
These electrons originate from nuclear decays from radon isotopes inside the spectrometer volume and can be magnetically trapped with high efficiency~\cite{PhDMertens2012,PhDWandkowsky2013}.
Secondary low-energetic electrons that are produced in scattering processes with residual gas can then leave the spectrometer and increase the observed background~\cite{Fraenkle2011,Mertens2013}.
The most effective and successful method to reduce this background is a passive technique where $\text{LN}_2$-cooled cryotraps inhibit the flow of radon molecules into the spectrometer volume, which significantly reduces the radon-induced background~\cite{Drexlin2017}.
Active methods target the remaining background by removing stored electrons from the MAC-E filter.
The `magnetic pulse method' decreases the strength of the magnetic guiding field\footnote{%
    This method was proposed by our collaborator E.~Otten, who suggested to break the magnetic storage conditions of trapped electrons inside the spectrometers for a short moment by a vanishing magnetic field.}
by inverting the air coil currents and thus widens the flux tube according to \eqref{eq:fluxtube}, which forces electrons towards the vessel walls where they are removed from the flux tube~\cite{PhDBehrens2016}.

\begin{figure}[tb]
    \center
    \includegraphics[height=4cm]{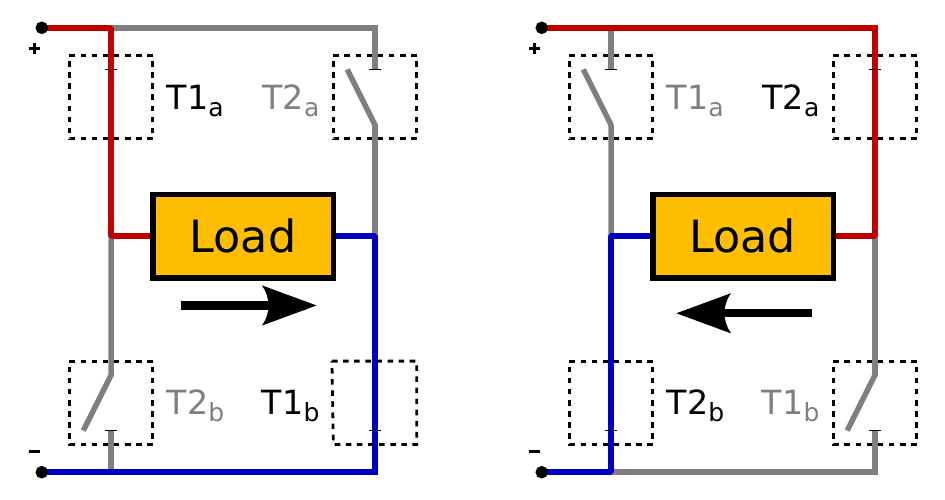}
    \quad
    \includegraphics[height=4cm]{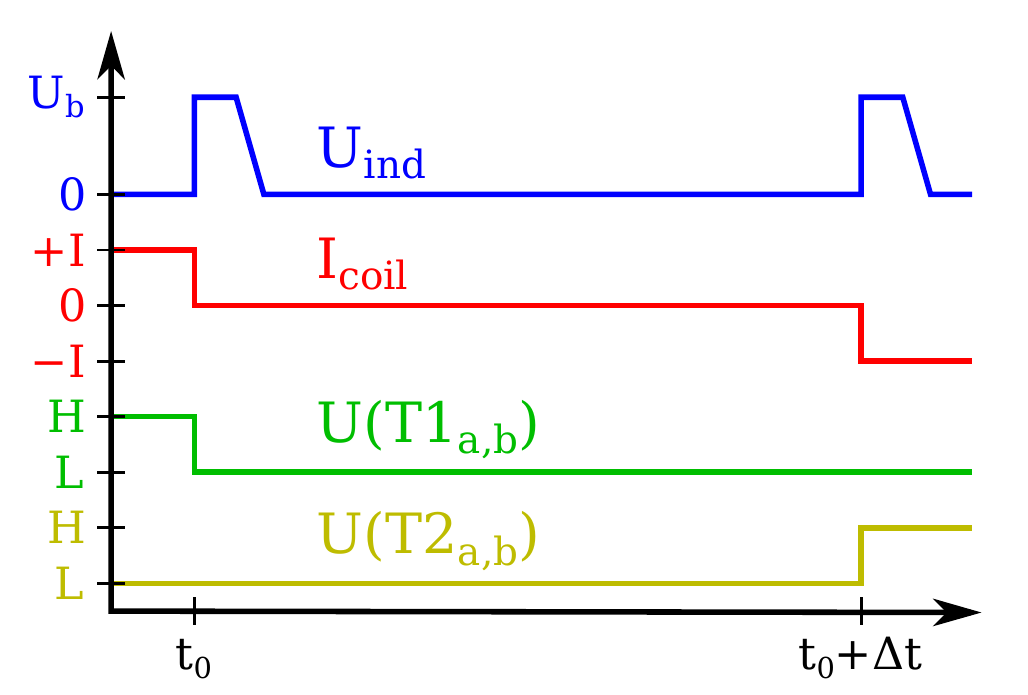}
    \caption{Concept of the H-bridge setup used in the flip-box units.
        Left: The circuit consists of four switches $\text{T1}_\mathrm{a,b}$, $\text{T2}_\mathrm{a,b}$ (MOSFETs) that are arranged in a H-like pattern.
        The polarity of the load circuit can be inverted without affecting the power supply.
        Right: The current direction $I_\mathrm{coil}$ in the load circuit is inverted by switching all four MOSFETs from high to low level and vice versa.
        The precise timing is established by a micro-processor that controls the gate voltage of the MOSFETs.
        A short delay ($\Delta t \approx \SI{10}{ms}$) is introduced to prevent over-shoot.
        Voltage spikes induced by the coil inductivity is limited to $U_\mathrm{b} = \SI{40}{V}$ by additional TVS diodes.}
    \label{fig:Hbridge}
\end{figure}

Fully inverting the air coil currents has the benefit that it only requires a polarity change in the coil circuit; the power supplies do not have to regulate to a new current set point.
One method to perform the current inversion without affecting the current source is a H-bridge setup (fig.~\ref{fig:Hbridge}, left).
The air coil polarities can be controlled remotely by providing a control signal for the micro-processor from the existing air coil slow-control system (see sec.~\ref{sec:sslowcontrol}).
The magnetic pulse system therefore allows to operate individual air coils with normal or inverted currents, both in dynamic or static inversion mode.

The so-called flip-box unit has been developed in~\cite{PhDBehrens2016} as a current-inversion device for the large-volume air coil system.
Each unit uses four high-current MOSFETs~\cite{PowerMOSFET} that support currents up to \SI{195}{A} at a maximum voltage of \SI{75}{V}.
It can invert the air coil polarity within \SI{10}{ms} during normal operation when currents are applied.
The inductance of the air coils induces a voltage spike when the load is disconnected from the power supply for a short time (see fig.~\ref{fig:Hbridge}, right), which can exceed the \SI{75}{V} limit of the MOSFETs.
Additional transient-voltage suppressor (TVS) diodes are embedded parallel to each MOSFET's source-drain junction to protect against over-voltage.
The diodes shorten the load circuit at voltages above $U_\mathrm{b} = \SI{40}{V}$ until the excess power is dissipated by the internal coil resistance, which is typically the case after a few \si{ms}.
To ensure stable operation at high currents, the MOSFETs are mounted on aluminum heat-sinks that are actively cooled by an air flow of \SI{46}{m^3/h} per unit; the air is taken in from the spectrometer hall with an ambient temperature of \SI{18}{\celsius}.
A total of 16 flip-box units were installed at the LFCS and EMCS coils and have been in continuous stable operation over several months during multiple measurements campaigns.
The devices were operated successfully at currents up to \SI{180}{A} and no problems with the electrical or thermal design were observed~\cite{PhDBehrens2016}.

\begin{figure}[tb]
    \center
    \includegraphics[width=0.80\textwidth]{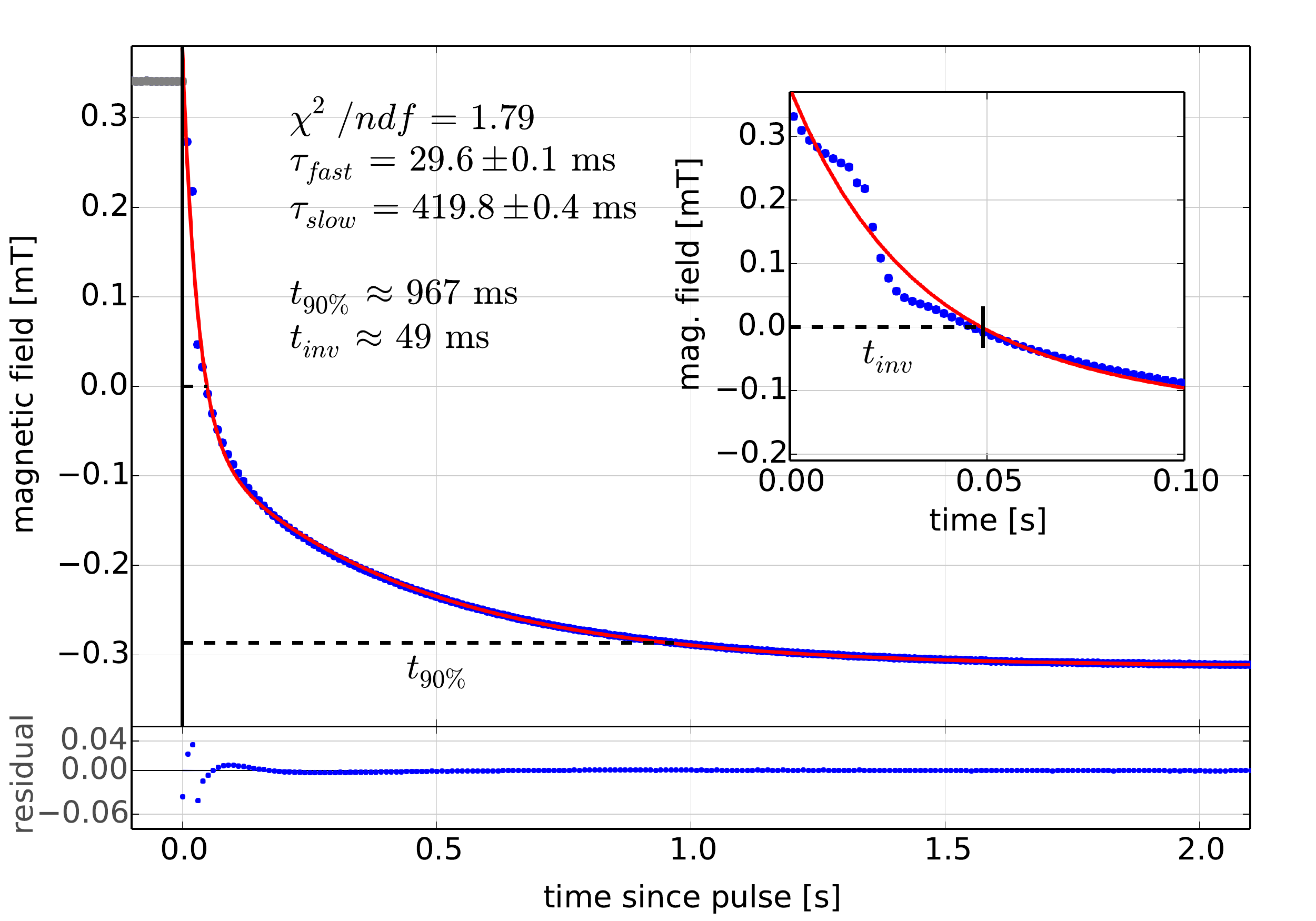}
    \caption{Magnetic field change at the spectrometer vessel during a magnetic pulse.
        The currents in LFCS air coils $\text{L1--L13}$ were inverted simultaneously at the \SI{0.38}{mT} setting and the magnetic field was measured with a Bartington-type sensor (interval \SI{10}{ms}, blue data points).
        The sensor was placed on the outer vessel wall close to air coil $\text{L8}$.
        The magnetic field change can be described by a double-exponential decay curve (red) with two independent time constants.
        The inset shows the field readings in the first \SI{0.2}{s}; the observed deviation is explained in the text.}
    \label{img:magpulse:magfield_bartington}
\end{figure}

Figure~\ref{img:magpulse:magfield_bartington} shows the magnetic field at the spectrometer vessel during a single magnetic pulse at a nominal field of \SI{0.38}{mT}.
The pulse was applied by inverting LFCS air coils $\text{L1--L13}$ simultaneously; coil $\text{L14}$ already has inverted polarity under normal conditions.
The field readings were performed with a fluxgate sensor~\cite{BartingtonMag03} that was placed manually at the outer vessel wall in vicinity of air coil $\text{L8}$ (close to the analyzing plane).

The field change by the magnetic pulse can be described by the sum of two exponential curves with time constants of $\tau_{slow} = \SI{419.8(4)}{ms}$ and $\tau_{fast} = \SI{29.6(1)}{ms}$~\cite{PhDBehrens2016}.
The fast time constant is assumed to be driven by the current inversion in the air coil, while the slow time constant corresponds to the overall magnetic field change, which is slowed down due to the mutual inductance between individual air coils and eddy current in the stainless steel vessel wall.
The inductance of $\text{L8}$ was estimated to $L \approx \SI{3}{mH}$ from an integration of the simulated magnetic flux over the coil's cross section. With the measured resistance of $R = \SI{0.144}{\Ohm}$, the time constant of $\tau = L/R \approx \SI{21}{ms}$ can be estimated.
The discrepancy to the measured time constant of $\SI{30}{ms}$ is explained by the inductive coupling between neighboring air coils, which increases the effective inductance. The power supplies have to regulate themselves on a short time-scale during the magnetic pulse due to the changed load when the H-bridge disconnects the air coil for a few \si{ms}. This explains the deviation from the theoretical (exponential) decay curve seen in the inset of fig.~\ref{img:magpulse:magfield_bartington}.
Since the deviations happen on a short time-scale that is much slower than the relaxation time of the magnetic field, they are not relevant for the operation of the magnetic pulse system. The slow time constant dominates the long-term behavior and defines the relaxation time $t_{90\%} \approx \SI{967}{ms}$, after which a stable inverted magnetic field is reached.
Measurements with an electron beam in the main spectrometer show that the magnetic field change inside the MAC-E filter is further slowed down by eddy currents~\cite{PhDBehrens2016}.
This increases the effective relaxation time and puts a lower limit of about \SI{1}{s} on the length of a magnetic pulse to achieve magnetic field inversion.

\section{Performance of the air coil system}
\label{SecPerformance}

As outlined in sec.~\ref{SecEMD}, the strength and shape of the magnetic field in the central region of the main spectrometer are crucial parameters for neutrino mass analysis.
The magnetic flux tube is mapped to the focal plane detector, which is segmented into 148 individual pixels~\cite{Amsbaugh2015}.
Each pixel thus corresponds to a specific section of the spectrometer volume in radius and azimuth.
The individual field $B_\mathrm{a}(r, \phi)$ in the analyzing plane must be reproduced by simulations with a precision of 1\% for a nominal magnetic field of \SI{0.33}{mT}, which achieves an energy resolution in the \si{eV}-regime for the MAC-E filter.

Direct field measurements inside the main spectrometer vessel are not possible since the installation of the inner electrode system has been completed~\cite{Valerius2010} and the spectrometer was commissioned for measurements at ultra-high vacuum.
Hence, information about the magnetic field distribution in the analyzing plane can only be obtained through detailed simulations.
An experimental validation of the magnetic field simulations for the air coil systems~\cite{Glueck2013} is of crucial importance for in-depth understanding of the properties of the MAC-E filter.

\subsection{Measurement procedure}
\label{MeasurementProcedure}

\begin{figure}[tb]
    \centering
    \includegraphics[width=0.90\textwidth]{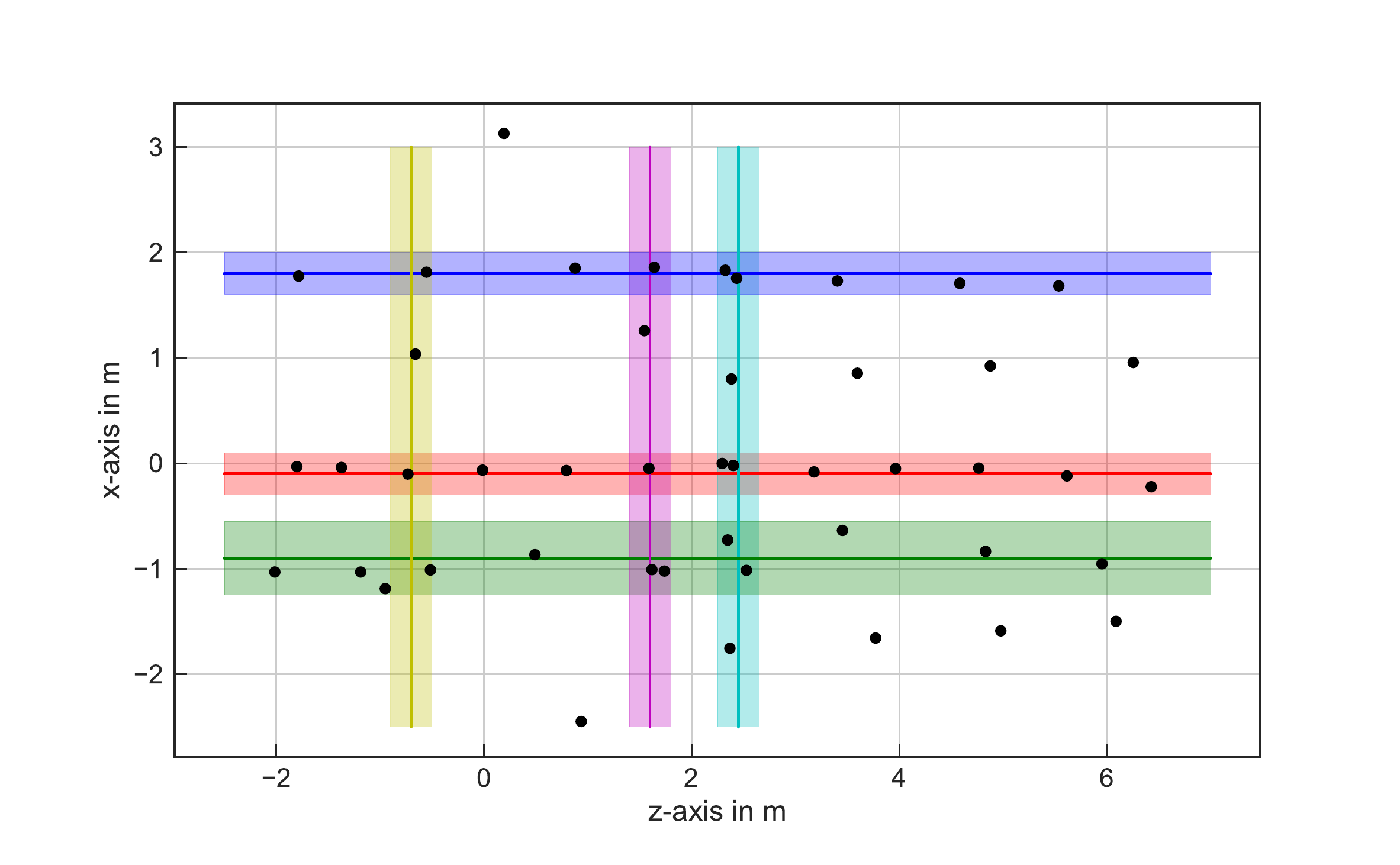}
    \caption{Positions of the magnetic field measurements inside the main spectrometer in the $x$-$z$-plane; the $y$-position is at \SI{-0.4}{m} for the entire data set.
        The point $x = y = 0$ corresponds to the KATRIN beam axis (symmetry axis of the main spectrometer) and the center of the main spectrometer is at $z = 0$; both positions are indicated by red lines.
        The colored bars indicate the data sets shown in fig.~\ref{fig:EMCS_along_r} and fig.~\ref{fig:LFCS_30}; these figures use the same respective colors as shown here.
    }
    \label{FigPath}
\end{figure}

To provide a basis for comparing the simulation results to measured magnetic fields, a series of field measurements inside the spectrometer was carried out during the installation of the inner-electrode system~\cite{PhDReich2013}.
These measurements were limited to the volume accessible by a scaffolding structure inside the vented spectrometer that was designed for the inner-electrode mounting procedure~\cite{Valerius2010,PhDZacher2015}.
Thereby a large-volume map of the magnetic field in the interior of the central main spectrometer section could be gained (fig.~\ref{FigPath}).

The magnetic field measurements employed three-axis fluxgate magnetometers of type Mag-03 from Bartington Instruments~\cite{BartingtonMag03}.
The sensors were mounted on an aluminum frame; embedded in this frame were four reflector spheres as reference points for a laser tracker system~\cite{LeicaTS30}.
In combination with an off-line calibration, this allows to determine the three-dimensional sensor orientation w.r.t. the spectrometer reference frame with an accuracy of $\le \ang{0.2}$.
The spatial resolution of the sensor position measurement has an accuracy of \SI{15}{mm}; the uncertainty is mainly due to the arrangement of the sensor elements within the magnetometer.
The sensors reach an accuracy of \SIrange{10}{20}{nT} for each field component and feature a stable offset of $\pm \SI{5}{nT}$ with a linearity of $< 0.0015\%$~\cite{BartingtonMag03}.
During the field measurements the air coil current set points were monitored with high precision DC/DC converters~\cite{DeltaIsoAmp}.
The maximum drift of the power supplies was measured to be $< \SI{20}{mA/month}$ over a time interval of about $\SI{100}{days}$~\cite{PhDErhard2016}.

With this setup, the magnetic field generated by the air coil system was measured at 46~positions in the spectrometer volume as shown in fig.~\ref{FigPath}.
All measurement points are distributed along the $x$--$z$ plane below the spectrometer axis at a fixed $y$-position of \SI{-0.40+-0.01}{m}.
At each position, the magnetic field was measured at a current of \SI{30}{A} for the individual LFCS coils.
The EMCS system was operated at \SI{50}{A} (vertical) and \SI{15}{A} (horizontal).
These values are selected since they correspond rather well to the nominal settings (see tab.~\ref{TabLFCS}) and make it easier to compare the contributions from individual air coils; investigations with different air coil settings can be found in~\cite{PhDReich2013,PhDErhard2016}.

In addition, the background magnetic field (a combination of the earth magnetic field and the remanent magnetic field of magnetized ferromagnetic materials surrounding the spectrometer) was measured at every position with zero air coil currents.
The background was then subtracted vectorially from the field measurements to obtain the contribution of the air coil systems to the global magnetic field.

\subsection{Magnetic field inside the spectrometer}

\begin{figure}[p]
    \centering
    \includegraphics[width=1.0\linewidth]{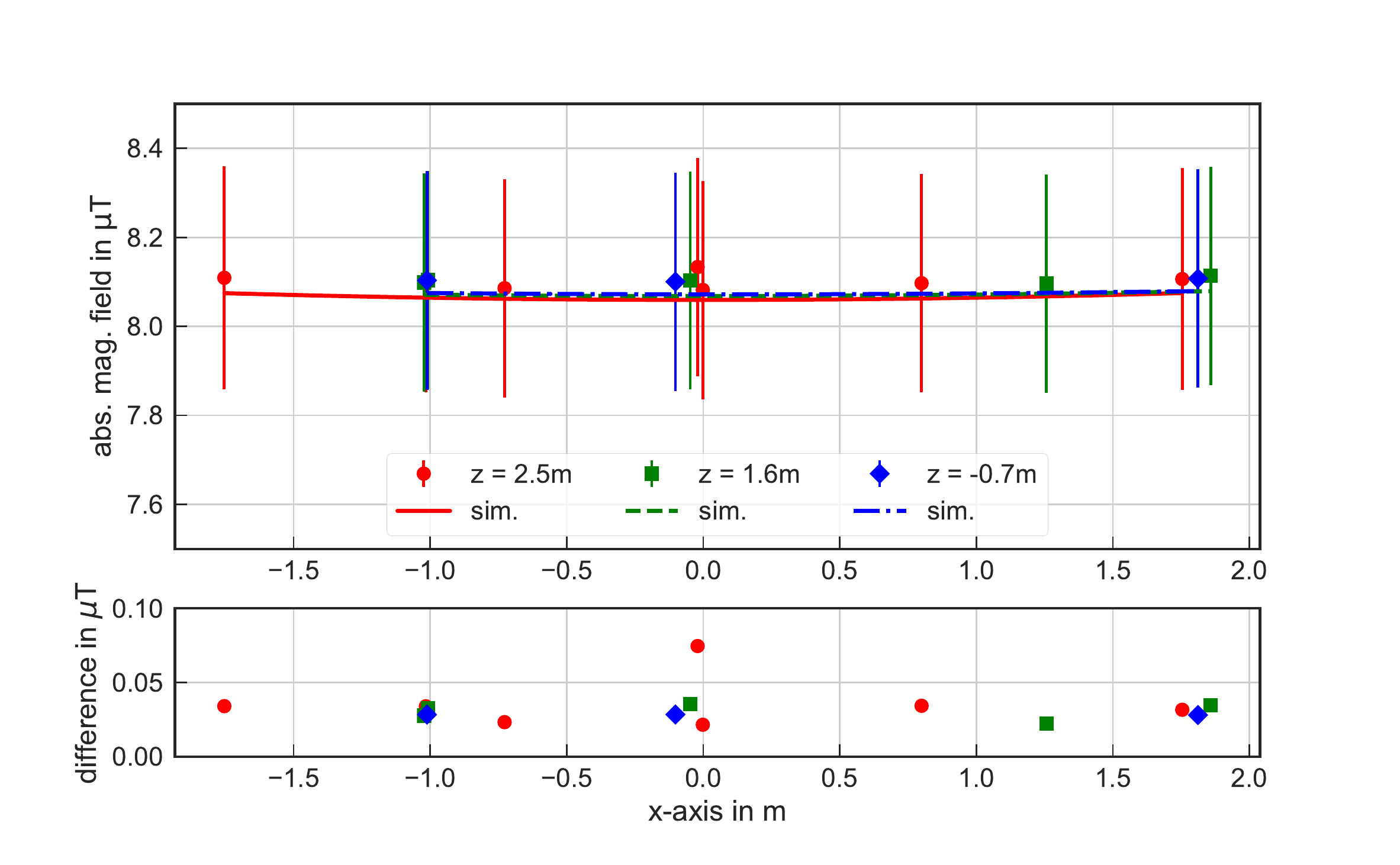}
    \\
    \includegraphics[width=1.0\linewidth]{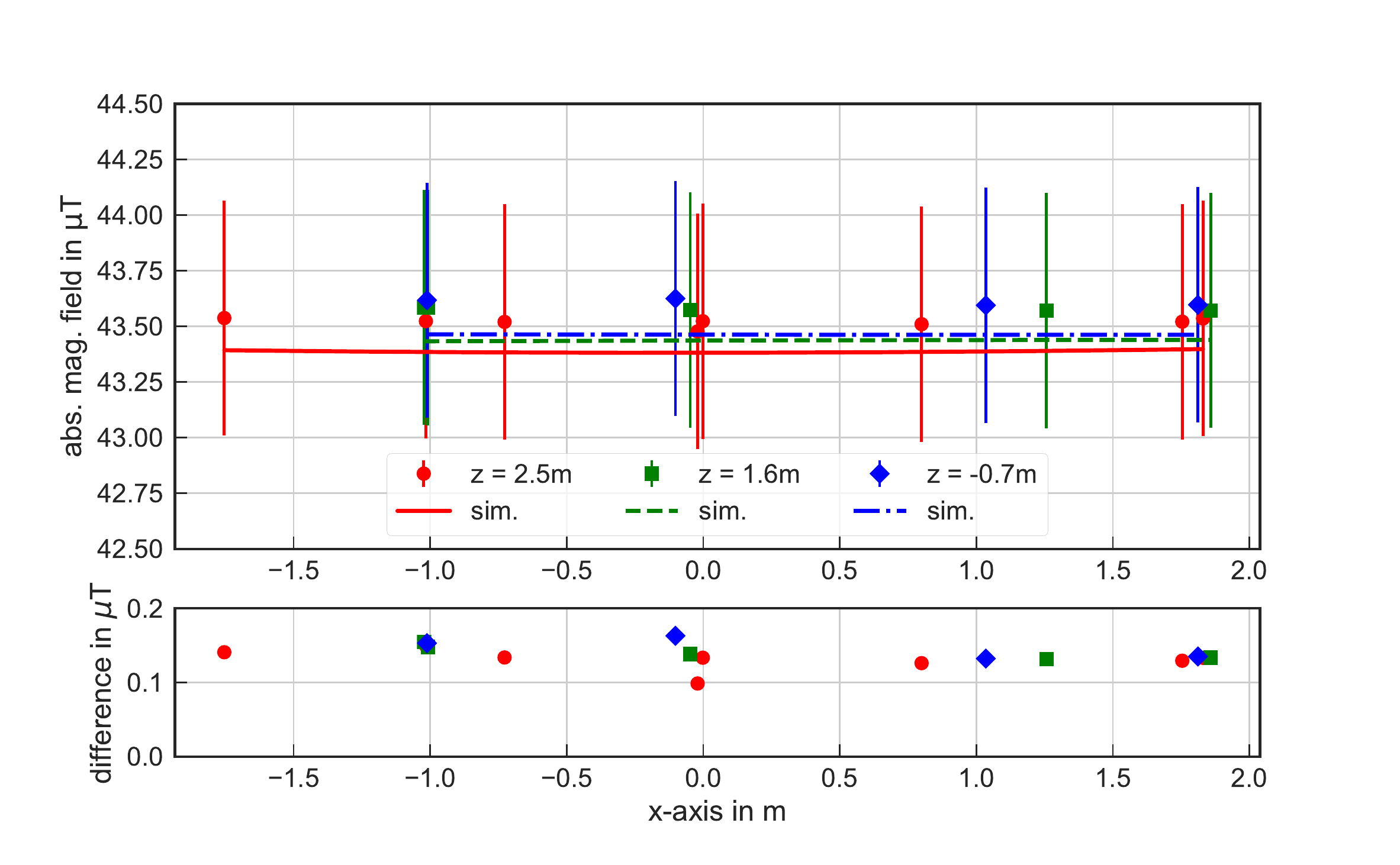}
    \caption{Comparison of the EMCS magnetic field at \SI{15}{A} (EMCS-X, upper panel) and \SI{50}{A} (EMCS-Y, lower panel) with simulations.
        The measurement positions in the main spectrometer are given along the radius at various axial positions (fig.~\ref{FigPath}).
        In both cases, the measurement confirms a homogeneous magnetic field within the given uncertainties.}
    \label{fig:EMCS_along_r}
\end{figure}

\begin{figure}[p]
    \centering
    \includegraphics[width=1.0\linewidth]{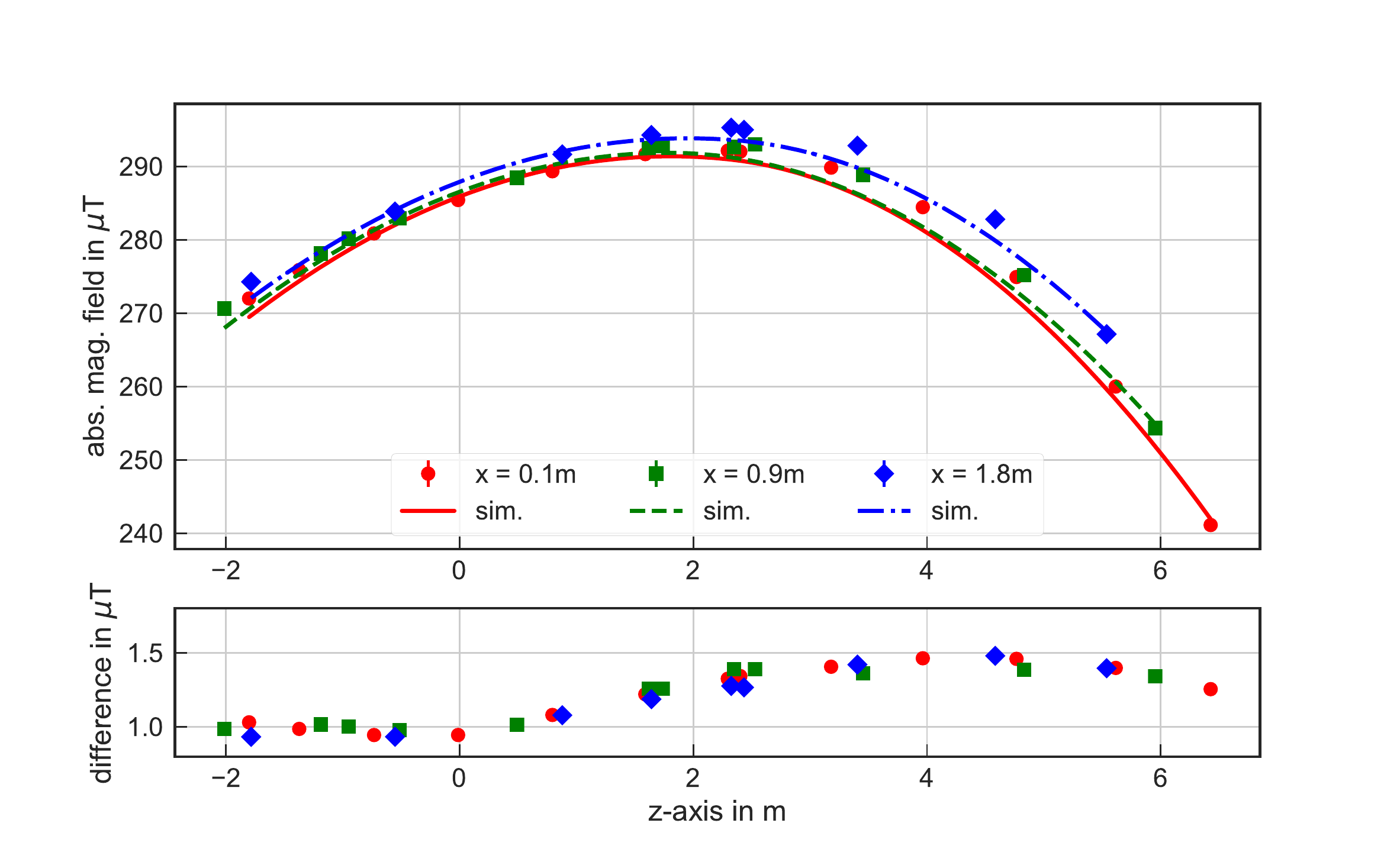}
    \\
    \includegraphics[width=1.0\linewidth]{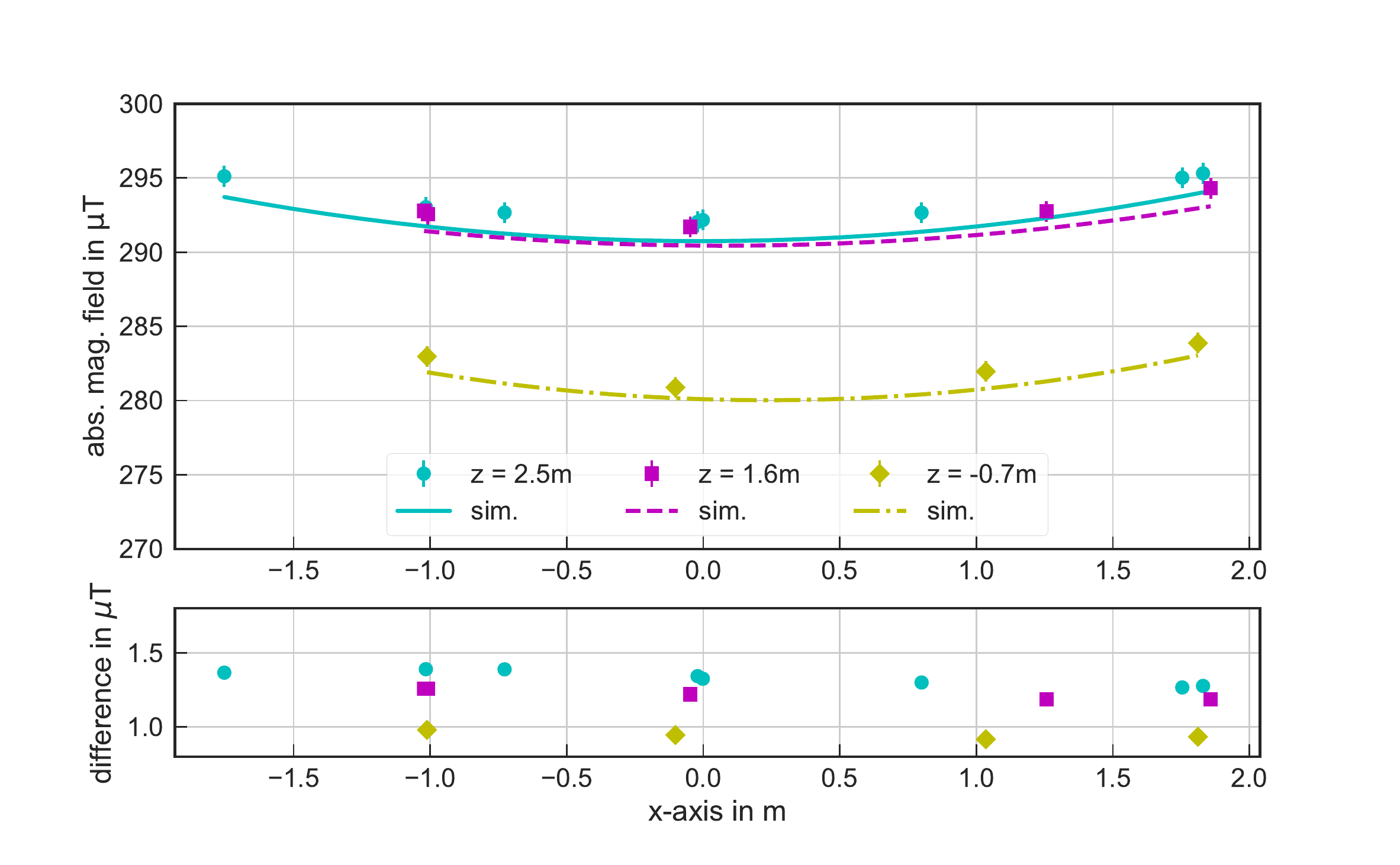}
    \caption{Comparison of the LFCS magnetic field at \SI{30}{A} for all coils with simulations.
        The measurement positions are along the spectrometer axis at various horizontal positions (upper panel) and axial positions (lower panel), as shown in fig.~\ref{FigPath}.
        The measurement uncertainties are approximately \SI{0.7}{\micro T}.
        Both plots include the magnetic field difference, where the axial measurement features a systematic increase towards larger $z$-values.
        Although the observed deviation is within the given uncertainty, it could be caused by induced magnetization of the steel in the spectrometer hall.
        All observed deviations to the field model are within the required \SI{2}{\micro T} limit.}
    \label{fig:LFCS_30}
\end{figure}

Fig.~\ref{fig:EMCS_along_r} shows the magnetic field generated by the EMCS system for the horizontal (vertical) field component in the upper (lower) panel.
For both sub-systems the measurement was performed in horizontal direction at three different axial positions ($z = \SIlist{-0.7;1.6;2.5}{m}$ with $y = \SI{-0.4}{m}$).
The field generated by the LFCS is shown in fig.~\ref{fig:LFCS_30} for a scan in axial (horizontal) direction in the upper (lower) panel.
In the first case the measurement was performed parallel to the spectrometer axis at three different radial positions ($x = \SIlist{-0.9;-0.1;1.8}{m}$ with $y = \SI{-0.4}{m}$); in the second case in the horizontal direction at three different axial positions ($z = \SIlist{-0.7;1.6;2.5}{m}$ with $y = \SI{-0.4}{m}$).
The colors used in these plots match the measurement point groups that are visualized in fig.~\ref{FigPath}.

The uncertainties of the magnetic field measurements incorporate the measured field fluctuations, the background field uncertainties and the uncertainty in sensor orientation.
For each measurement, a field simulation was performed with \textsc{Kassiopeia} (sec.~\ref{sec:KASSIOPEIA}).
The simulation is based on the realistic geometry of the air-coil system (sec.~\ref{sec:mechanical}) and uses the currents measured at the power supplies by current transducers (sec.~\ref{sec:electrical} and~\ref{sec:sslowcontrol}).
The deviation of the absolute fields is then calculated as $B_\mathrm{diff} = |\vec{B}_\mathrm{meas} - \vec{B}_\mathrm{sim}|$.

A good agreement of the field modeling to the experimental data is achieved in all test cases as shown in fig.~\ref{fig:EMCS_along_r} and \ref{fig:LFCS_30}, satisfying the key requirement that all field deviations are below \SI{2}{\micro T}~\cite{PhDGroh2015,PhDErhard2016}.
The influence of the magnetic field uncertainty in the analyzing plane on the neutrino mass systematic of the KATRIN experiment is therefore negligible under nominal conditions.

In the case of the EMCS a complete agreement within the calculated uncertainties is observed.
This test also verifies that the EMCS works as intended and provides a homogeneous field inside the spectrometer vessel in both field directions.
The remaining field deviations are in the sub-\si{\micro T} regime, and are thus of the same order as the uncertainties introduced by current fluctuations and deviations from the current set-points (\cite{PhDErhard2016}).
The observed deviations in the field model introduced by the EMCS are therefore of no concern for the KATRIN experiment.

The test of the LFCS shows also that it works as intended and generates an almost homogeneous field near the analyzing plane at $z = 0$.
Field variations of a few \si{\micro T} are observed in radial direction; in the axial range of $z = \pm \SI{2}{m}$ in the central spectrometer region, the maximum field difference is less than \SI{1.5}{\micro T}.
The absolute difference between measurement and simulations is still below \SI{2}{\micro T} for any measurement point.
The air coil system is therefore ready for operation in a neutrino mass measurement campaign at nominal magnetic field.

Interestingly, a small systematic offset is observed for all field measurements, which we will discuss shortly.
One must consider that any inaccuracy in geometric data (simulation model) or in the current readings is directly translated to an error of the magnetic field simulation; the impact of inaccurate currents is further multiplied by the number of coil windings (sec.~\ref{SecTD}).
However, due to the accurately determined geometry of the air coil systems (fig.~\ref{fig:grav_deform}) and the precise current readouts (sec.~\ref{MeasurementProcedure}), these parameters can be excluded as the origin of the observed deviation.
The remanent magnetic field of the ferromagnetic structural materials surrounding the spectrometer was already subtracted from these measurements.
Hence, the only viable explanation for the observed offset is an \emph{induced} magnetic field in the steel rods of the reinforced concrete walls of the spectrometer hall (see fig.~\ref{FigBuilding}).
    
For standard operation of the KATRIN experiment with nominal magnetic fields around \SI{0.3}{mT}, the induced magnetization is of no concern as its field contribution in the flux tube is below the required accuracy.
However, for larger magnetic fields above \SI{0.4}{mT} this effect becomes not only more profound, but also has a stronger impact on the experimental sensitivity~\cite{PhDErhard2016}.
This issue must be addressed if increased magnetic fields are applied during neutrino mass measurements, e.g. to reduce the observed background and thus improve the neutrino mass sensitivity by reducing the flux tube volume~\cite{Fraenkle2017}.

One can define a simplified model of the magnetic materials by placing a small number of dipole moments around the main spectrometer volume.
The strength of the dipole moments is then fitted to magnetic field measurements from the RMMS/MobSU system, which yields a contribution on the order of $< \SI{5}{\micro T}$ for the remanent magnetic field~\cite{PhDErhard2016}.
An improvement can be achieved by defining a larger number of dipole moments, which requires a larger set of measurement data to prevent over-fitting.
A corresponding measurement that provides the necessary data will use the RMMS and VMMS field sensors and is planned for the near future.

\subsection{Magnetic field outside the spectrometer and air coil linearity}

A key assumption in the operation of the air coil system is that the magnetic field within the main spectrometer depends linearly on the currents applied to the air coil system.
Since the magnetic fields produced by the air coil systems are not particularly strong, the linear relation between current and magnetic field is expected to be straight-forward.
However, the air coil system encloses a large volume and could influence surrounding magnetic materials in the spectrometer hall, which could result in deviations from this linear behavior.
To investigate this possibility, we measured the magnetic field using different coils at varying current setpoints of the air coil power supplies.
This measurement was carried out after the installation of the wire electrode and is thus unrelated to the field measurements discussed earlier in this section.

\begin{figure}[htb]
    \centering
    \includegraphics[width=0.90\textwidth]{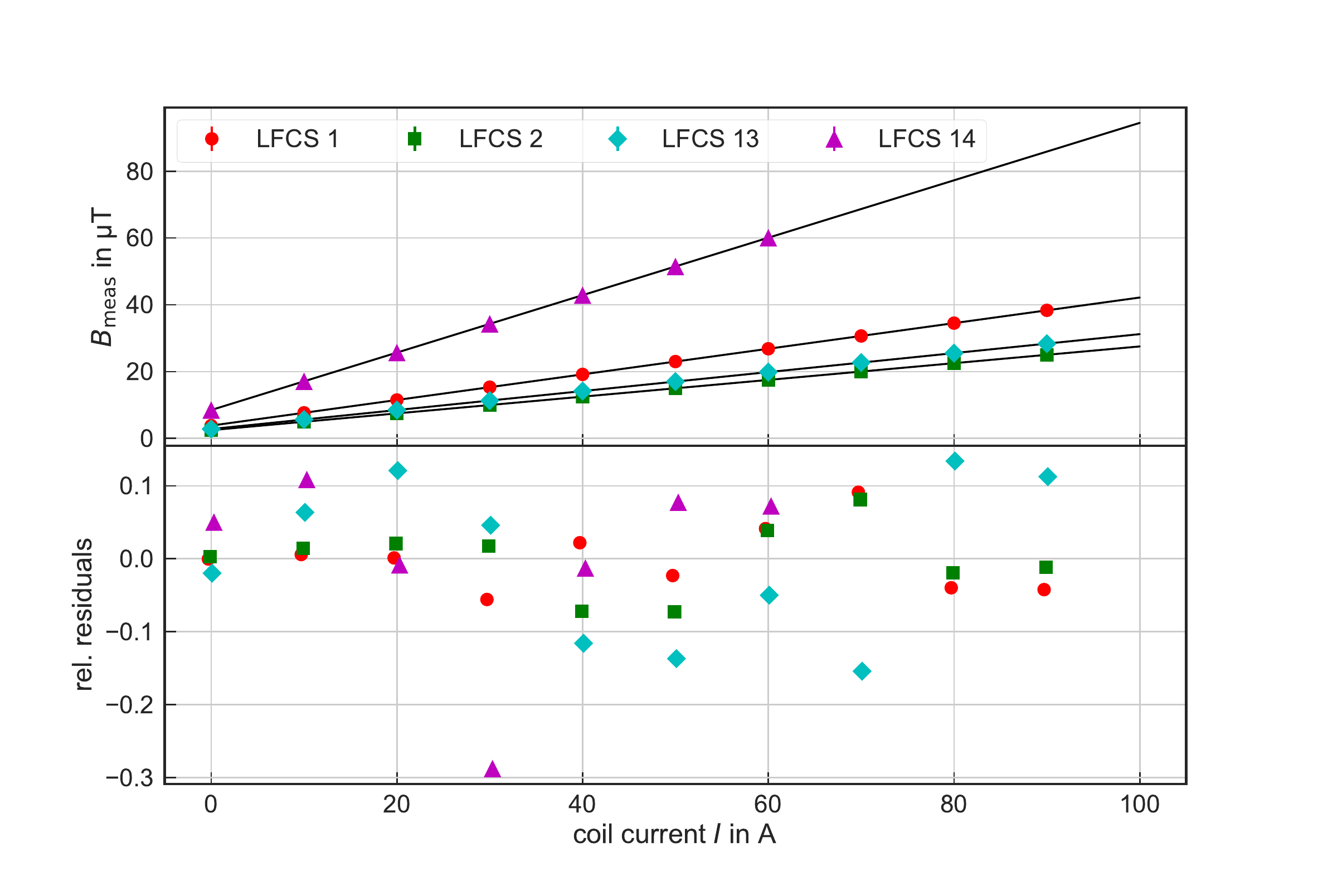}
    \caption{Measured magnetic field values for varying LFCS currents. This figure uses the current setpoints of the corresponding power supplies to investigate the response of the air coil system to varying operational parameters.}
    \label{Figlinearity}
\end{figure}

Fig.~\ref{Figlinearity} shows the magnetic field produced by four individual LFCS air coils at different currents.
We use the current setpoints here to determine the response of the air coil system to the applied operational parameters.
The magnetic fields were measured at a precisely known location outside of the spectrometer vessel.
For air coils $\text{L1+L2}$ the sensor position was in the southern flat cone section of the spectrometer (negative $z$-value, see fig.~\ref{Fig:BSettingWithPuls} and \ref{fig:LFCS}), and likewise for air coils $\text{L13+L14}$ in the northern flat cone section (positive $z$-value).
The data points in each measurement are fitted with a linear function, using an estimated measurement error of \SI{55}{nT} at each data point.
The bottom panel of fig.~\ref{Figlinearity} shows the normalized residuals.
No deviation from linearity to the applied current is observed in the four test cases and there is no visible trend in the residuals.

The small dependency of the residuals on the coil current hints at contributions from induced magnetic fields by the surrounding structural materials in the main spectrometer hall.
To completely assess the impact of magnetic materials one must also consider the operation of the superconducting solenoids since they produce stronger stray fields in their immediate vicinity and could thus increase the magnetization.
The VMMS (sec.~\ref{sec:monitoring}) will help to investigate the influence of such materials in the near future.

\section{Conclusion}
\label{SecConclusion}

The KATRIN experiment aims to determine the effective electron neutrino mass down to \SI{0.2}{eV} (90\%~C.L.) by measuring the integral electron energy spectrum close to the endpoint of molecular tritium beta-decay.
The precision energy filtering of the signal electrons takes place in the main spectrometer that features a small magnetic field $< \SI{0.5}{mT}$ at its center.
The task of the large-volume air coil system described in this paper is to fine-tune the magnetic guiding field provided by superconducting solenoids and to compensate the earth magnetic field inside the main spectrometer.

The LFCS (Low Field Correction System) allows to tune the axisymmetric magnetic field inside the main spectrometer vessel to the desired shape and strength; the field at the spectrometer center can be scaled up to the maximum value of \SI{1.8}{mT}.
The LFCS consists of 14 air coils with \SI{12.6}{m} diameter that are arranged coaxially with the main spectrometer vessel and the adjacent superconducting solenoids.
The EMCS (Earth Magnetic field Compensation System) compensates the earth magnetic field inside the main spectrometer.
It consists of two cosine coil systems, oriented in vertical and horizontal direction around the spectrometer.
The vertical EMCS has 16 current loops with horizontal planes to compensate the vertical earth magnetic field component (\SI{43.7}{\micro T}) at a current of about \SI{50}{A}; the horizontal EMCS has 10 current loops with vertical planes to compensate the horizontal earth magnetic field component which is perpendicular to the beam axis (\SI{4.9}{\micro T}) at a current of about \SI{9}{A}.
Each current loop consists of two linear sections parallel to the beam axis that are connected at two endrings.
The support structure of the air coil system consists of 25 aluminum rings, thus providing a low overall weight and optimal magnetic properties.
The linear segments of the EMCS cables are fixed inside aluminum pipes that are axially mounted on the support structure.
Both air coil systems use aluminum cables with \SI{70}{mm^2} cross section and \SI{1.5}{mm} polyethylene insulation.
The 14~LFCS and 2~EMCS coils are powered by individual power supplies that operate at currents up to \SI{175}{A} (see tab.~\ref{TabLFCS}).
Dedicated current-inverter units allow to invert the polarity of the individual coils without any hardware changes.

It is not possible to perform direct magnetic field measurements inside the main spectrometer vessel after the installation of the wire electrode system.
Therefore, we carried out measurements with different air coil configurations at various positions inside the vessel during the installation of the inner-electrode system.
We compared the measured values with simulations using our \textsc{Kassiopeia} software.
The magnetic field was measured with fluxgate sensors at given positions and orientations w.r.t. the spectrometer, both with zero air coil currents and with close-to-nominal air coil current setting.
The vectorial difference of the two measured values was then compared with simulation results.
The relative deviation between the measured and simulated values is about 1\% or smaller for the LFCS (0.4\% or smaller for the EMCS).
In addition, the linearity between the magnetic field and the coil currents was verified by measurements at several positions outside the spectrometer vessel.
The good agreement between our field measurements and simulations ensures that the magnetic field of the air coil system inside the main spectrometer vessel can be computed with a relative accuracy better than 1\%.
This fulfills the KATRIN specifications for a \SI{0.3}{mT} magnetic field in the main spectrometer, which corresponds to a \SI{1}{eV} energy resolution for \SI{18.6}{keV} electrons and is the design value for the neutrino mass measurement campaign.
An indirect measurement of the magnetic field in the spectrometer and hence a verification of the magnetic field model is possible with a dedicated electron source.
The emitted electrons are then used to probe the electromagnetic field in the spectrometer with high accuracy and spatial resolution.

In summary, we have shown that both air coil systems at the KATRIN main spectrometer are fully operational and that their contributions to the guiding magnetic field can be described with high accuracy on the sub-percent level by magnetic field simulations.
This is an important result for a detailed modeling of the transmission and background properties of the MAC-E filter setup at the main spectrometer, which is a key requirement for future neutrino mass analysis at the KATRIN experiment.

\acknowledgments
We acknowledge the support by H.~Hucker (KIT) in designing the air coil system, J.~Grimm (KIT) in carrying out the resisitivity measurements and K.~Mehret (KIT) for providing photographs of the air coil systems.
This work has been supported by the Bundesministerium f{\"u}r Bildung und Forschung (BMBF) with project numbers 05A17PM3 and 05A17VK2 and the Helmholtz Association (HGF).
We also would like to thank the Karlsruhe House of Young Scientists (KHYS) of KIT for their support (B.~L., S.~M., J.~R., N.~W.).
\paragraph{Disclaimer:}
Certain commercial equipment, instruments or materials are identified in this paper in order to adequately specify the environmental and experimental procedures. Such identification does not imply recommendation or endorsement by us, nor does it imply that the materials or equipment identified are necessarily the best available for the purpose.
%
%
%
\bibliography{RefAircoils,RefArXivPublications,RefGeneralPublications,RefKATRINDiplomaMasterTheses,RefKATRINPhDTheses,RefKATRINPublications}
\bibliographystyle{JHEP}
\end{document}